\documentclass[12pt]{article}
\pdfoutput=1
\usepackage{jheppub}
\usepackage{epsfig}
\usepackage{amsmath}
\usepackage{amssymb}
\usepackage{amsfonts}
\usepackage{amsxtra}
\usepackage{amsthm}
\usepackage{mathrsfs}
\usepackage{makeidx}
\usepackage{graphics}
\usepackage{dsfont}
\usepackage{mathtools}
\usepackage{graphicx}
\usepackage{placeins}
\usepackage{bm}
\usepackage[capitalise]{cleveref}
\usepackage{empheq}
\usepackage{colortbl}
\usepackage{xcolor}
\usepackage{enumerate}
\usepackage{titlesec}
\usepackage{longtable}
\usepackage{float}
\usepackage{color}
\usepackage{tikz}
\usepackage{xfrac}
\usepackage{fnpct}
\usepackage{footnote}
\usepackage{rotating}
\usepackage{lscape}
\usepackage{makecell}
\usepackage{environ}
\usepackage{tabularx}
\usepackage{subfiles}
\usepackage[export]{adjustbox}
\usepackage{ytableau}
\usepackage{tikz-3dplot}
\usepackage{slashed}
\usepackage{pifont}
\usepackage{multirow}
\usepackage{mdframed}
\usepackage{bbm}
\usepackage[normalem]{ulem}
\usepackage{enumitem}

\usetikzlibrary{positioning,trees,decorations.pathmorphing,decorations.markings,decorations.pathreplacing,calc,shapes,patterns,arrows,chains,arrows.meta,fit,fadings,decorations.markings,graphs,graphs.standard,quotes,plotmarks}

\DeclareMathOperator{\U}{U}
\DeclareMathOperator{\SU}{SU}
\DeclareMathOperator{\SO}{SO}

\DeclareMathOperator{\USp}{USp}

\newcommand{\coma}{\, , \quad}
\newcommand{\fstop}{\, .}

\def\ZZ{{\mathbb{Z}}}

\theoremstyle{plain}% default
\theoremstyle{definition}
\newtheorem{example}{Example}%[section]

\theoremstyle{remark}

\catcode`\@=11   
\newdimen\@rotdimen
\newbox\@rotbox  

\def\@vspec#1{\special{ps:#1}}%  passes #1 verbatim to the output
\def\@rotstart#1{\@vspec{gsave currentpoint currentpoint translate
   #1 neg exch neg exch translate}}% #1 can be any origin-fixing transformation
\def\@rotfinish{\@vspec{currentpoint grestore moveto}}% gets back in synch 
\def\@rotr#1{\@rotdimen=\ht#1\advance\@rotdimen by\dp#1%
   \hbox to\@rotdimen{\hskip\ht#1\vbox to\wd#1{\@rotstart{90 rotate}%
   \box#1\vss}\hss}\@rotfinish}
\def\@rotl#1{\@rotdimen=\ht#1\advance\@rotdimen by\dp#1%
   \hbox to\@rotdimen{\vbox to\wd#1{\vskip\wd#1\@rotstart{270 rotate}%
   \box#1\vss}\hss}\@rotfinish}%
\def\@rotu#1{\@rotdimen=\ht#1\advance\@rotdimen by\dp#1%
   \hbox to\wd#1{\hskip\wd#1\vbox to\@rotdimen{\vskip\@rotdimen
   \@rotstart{-1 dup scale}\box#1\vss}\hss}\@rotfinish}%
\def\@rotf#1{\hbox to\wd#1{\hskip\wd#1\@rotstart{-1 1 scale}%
   \box#1\hss}\@rotfinish}%
\def\rotate{\@ifnextchar[{\@rotate}{\@rotate[l\right]}}
\def\@rotate[#1]#2{\setbox\@rotbox=\hbox{#2}\@nameuse{@rot#1}\@rotbox}

\catcode`\@=12
\usetikzlibrary{positioning}
\usetikzlibrary{chains}
\usetikzlibrary{arrows, arrows.meta ,fit,decorations.pathreplacing}
\tikzstyle{every picture}+=[remember picture]
\tikzstyle{na} = [baseline]
\tikzstyle{ligne}=[draw, thick]

\usetikzlibrary{arrows, decorations.markings, calc, fadings, decorations.pathreplacing, patterns, decorations.pathmorphing, positioning}
\tikzset{>={Latex[width=1.5mm,length=1.5mm]}}
\tikzset{bd/.style={circle, draw=black, inner sep=0pt, fill=black, minimum size=1.2mm}}
\tikzset{bld/.style={circle, draw=blue, inner sep=0pt, fill=blue, minimum size=1.2mm}}
\tikzset{wd/.style={circle, draw=black, inner sep=0pt, fill=white, minimum size=1.2mm}}
\tikzset{rd/.style={circle, draw=red, inner sep=0pt, fill=red, minimum size=.9mm}}
\tikzset{wrd/.style={circle, draw=red, inner sep=0pt, fill=white, minimum size=.9mm}}
\usetikzlibrary{graphs,graphs.standard,quotes}

\def\node#1#2{\overset{#1}{\underset{#2}{{\color{gray} \bullet}}}}

\def\node#1#2{\overset{#1}{\underset{#2}{\circ}}}

\tikzstyle{every picture}+=[remember picture]
\tikzstyle{na} = [baseline=-.5ex]

\newcommand{\eg}{e.g. }

\newcommand{\ie}{i.e. }

\numberwithin{equation}{section}
\newcommand{\bes}[1]{\begin{equation} \begin{split} #1\end{split} \end{equation}}

\newcommand{\nn}{\nonumber}

\newcommand{\be}{\begin{equation}} \newcommand{\ee}{\end{equation}}
\newcommand{\bea}{\begin{equation} \begin{aligned}} \newcommand{\eea}{\end{aligned} \end{equation}}

\def\tilde{\widetilde}

\def\hat{\widehat}

\def\rt2{\sqrt{2}}

\def\CE{{\cal E}}

\def\CH{{\cal H}}
\def\CI{{\cal I}}

\def\CN{{\cal N}}

\def\CP{{\cal P}}

\def\CS{{\cal S}}
\def\CT{{\cal T}}

\def\CW{{\cal W}}

\def\CZ{{\cal Z}}

\def\1{{\ds 1}}

\newcommand{\cE}{\mathcal{E}}

\newcommand{\cZ}{\mathcal{Z}}

\newcommand{\fm}{\mathfrak{m}}

\def\SO{\mathrm{SO}}
\def\O{\mathrm{O}}
\def\SU{\mathrm{SU}}

\def\Spin{\mathrm{Spin}}

\def\su{\mathfrak{su}}

\def\so{\mathfrak{so}}

\def\usp{\mathfrak{usp}}
\def\fp{\mathfrak{p}}

\def\fN{\mathfrak{N}}

\def\repa{\raise4pt\hbox{$\square$}\mkern-14mu\raise-4pt\hbox{$\square$}}
\def\repab{\overline{\raise4pt\hbox{$\square$}\mkern-14mu\raise-4pt\hbox{$\square$}\mkern-1mu}}

\def\smileface{\ensuremath{\hbox{\large$\bigcirc$}\mkern-15mu\raise-1pt\hbox{\scriptsize$\smallsmile$}%
\mkern-10mu\raise4pt\hbox{..}\mkern4mu}}
\def\frownface{\ensuremath{\hbox{\large$\bigcirc$}\mkern-15mu\raise-1pt\hbox{\scriptsize$\smallfrown$}%
\mkern-10mu\raise4pt\hbox{..}\mkern4mu}}

\newcommand{\ba}{\begin{array}}
\newcommand{\ea}{\end{array}}
\newcommand{\bi}{\begin{itemize}}
\newcommand{\ei}{\end{itemize}}

\def\bea#1\eea{\allowdisplaybreaks \begin{align}#1\end{align}}
 \newcommand{\ben}{\begin{enumerate}}
\newcommand{\een}{\end{enumerate}}
\newcommand{\bean}{\begin{eqnarray*}}
\newcommand{\eean}{\end{eqnarray*}}
\newcommand{\eref}[1]{(\ref{#1})}

\newcommand{\PE}{\mathop{\rm PE}}

\newcommand{\BC}{\mathbb{C}}

\newcommand{\BZ}{\mathbb{Z}}

\definecolor{light-gray}{gray}{0.5}

\def\aup#1 {\overset{#1}{\uparrow} \, \overset{\tilde{#1}}{\downarrow}}

\tikzset{snake it/.style={decorate, decoration={snake, amplitude=.4mm, segment length=2mm,
                       post length=0mm,pre length=0mm}}}
                       
 \newcommand{\GCD}{\mathrm{GCD}}

\hypersetup{
	pdftitle={A tale of 2-groups: \texorpdfstring{$D_p(\USp(2N))$}{} theories},    % title
	pdfauthor={\textcopyright\ Federico Carta, Simone Giacomelli, Noppadol Mekareeya, Alessandro Mininno},     % author
	pdfsubject={HEP},   % subject of the document
	pdfcreator={pdfLaTex},   % creator of the document
	pdfproducer={LaTex}, % producer of the document
	pdfkeywords={},
	colorlinks=true,
}

\makeatletter
\newsavebox{\measure@tikzpicture}
\NewEnviron{scaletikzpicturetowidth}[1]{%
  \def\tikz@width{#1}%
  \begin{lrbox}{\measure@tikzpicture}%
  \BODY
  \end{lrbox}%
  \pgfmathparse{#1/\wd\measure@tikzpicture}%
  \BODY
}
\makeatother

\makeatletter
\def\squarecorner#1{
    \pgf@x=\the\wd\pgfnodeparttextbox%
    \pgfmathsetlength\pgf@xc{\pgfkeysvalueof{/pgf/inner xsep}}%
    \advance\pgf@x by 2\pgf@xc%
    \pgfmathsetlength\pgf@xb{\pgfkeysvalueof{/pgf/minimum width}}%
    \ifdim\pgf@x<\pgf@xb%
        \pgf@x=\pgf@xb%
    \fi%
    \pgf@y=\ht\pgfnodeparttextbox%
    \advance\pgf@y by\dp\pgfnodeparttextbox%
    \pgfmathsetlength\pgf@yc{\pgfkeysvalueof{/pgf/inner ysep}}%
    \advance\pgf@y by 2\pgf@yc%
    \pgfmathsetlength\pgf@yb{\pgfkeysvalueof{/pgf/minimum height}}%
    \ifdim\pgf@y<\pgf@yb%
        \pgf@y=\pgf@yb%
    \fi%
    \ifdim\pgf@x<\pgf@y%
        \pgf@x=\pgf@y%
    \else
        \pgf@y=\pgf@x%
    \fi
    \pgf@x=#1.5\pgf@x%
    \advance\pgf@x by.5\wd\pgfnodeparttextbox%
    \pgfmathsetlength\pgf@xa{\pgfkeysvalueof{/pgf/outer xsep}}%
    \advance\pgf@x by#1\pgf@xa%
    \pgf@y=#1.5\pgf@y%
    \advance\pgf@y by-.5\dp\pgfnodeparttextbox%
    \advance\pgf@y by.5\ht\pgfnodeparttextbox%
    \pgfmathsetlength\pgf@ya{\pgfkeysvalueof{/pgf/outer ysep}}%
    \advance\pgf@y by#1\pgf@ya%
}
\makeatother

\pgfdeclareshape{square}{
    \savedanchor\northeast{\squarecorner{}}
    \savedanchor\southwest{\squarecorner{-}}

    \foreach \x in {east,west} \foreach \y in {north,mid,base,south} {
        \inheritanchor[from=rectangle]{\y\space\x}
    }
    \foreach \x in {east,west,north,mid,base,south,center,text} {
        \inheritanchor[from=rectangle]{\x}
    }
    \inheritanchorborder[from=rectangle]
    \inheritbackgroundpath[from=rectangle]
}

\tikzset{stretch/.initial=1}
\newcommand\drawloop[4][]%
   {\draw[shorten <=0pt, shorten >=0pt,#1]
      ($(#2)!\pgfkeysvalueof{/tikz/stretch}!(#2.#3)$)
      let \p1=($(#2.center)!\pgfkeysvalueof{/tikz/stretch}!(#2.north)-(#2)$),
          \n1= {veclen(\x1,\y1)*sin(0.5*(#4-#3))/sin(0.5*(180-#4+#3))}
      in arc [start angle={#3-90}, end angle={#4+90}, radius=\n1]%
   }

\preprint{ZMP-HH/22-16}
\title{A tale of 2-groups: \texorpdfstring{D$_{\bm{p}}$(\text{USp}(2N))}{} theories}
\author[a]{Federico Carta,}
\author[b]{~Simone Giacomelli,}
\author[c,d,e]{~Noppadol Mekareeya}
\author[f]{\\ and Alessandro Mininno}
\affiliation[a]{Department of Mathematical Sciences,
		Durham University, \\ Durham, DH1 3LE, United Kingdom}
\affiliation[b]{Dipartimento di Fisica, Universit\`a di Milano-Bicocca, Piazza della Scienza 3, \\ I-20126 Milano, Italy}
\affiliation[c]{INFN, sezione di Milano-Bicocca, Piazza della Scienza 3, \\ I-20126 Milano, Italy}
\affiliation[d]{Laboratoire de Physique de l'\'Ecole Normale Sup\'erieure, CNRS, \\  Universit\'e PSL, Sorbonne Universit\'e, 24 rue Lhomond, 75005 Paris, France}
\affiliation[e]{Department of Physics, Faculty of Science, Chulalongkorn University, \\ Phayathai Road,
Pathumwan, Bangkok 10330, Thailand}
\affiliation[f]{II. Institut f\"ur Theoretische Physik, Universit\"at Hamburg,\\
Luruper Chaussee 149, 22607 Hamburg, Germany}
\emailAdd{federico.carta@durham.ac.uk}
\emailAdd{simone.giacomelli@unimib.it}
\emailAdd{n.mekareeya@gmail.com}
\emailAdd{alessandro.mininno@desy.de}
\abstract{A 1-form symmetry and a 0-form symmetry may combine to form an extension known as the 2-group symmetry. We find the presence of the latter in a class of Argyres-Douglas theories, called $D_p($USp$(2N))$, which can be realized by $\mathbb{Z}_2$-twisted compactification of the 6d $\mathcal{N}=(2,0)$ of the $D$-type on a sphere with an irregular twisted puncture and a regular twisted full puncture. We propose the $3$d mirror theories of general $D_p($USp$(2N))$ theories that serve as an important tool to study their flavor symmetry and Higgs branch. Yet another important result is presented: We elucidate a technique, dubbed ``bootstrap'', which generates an infinite family of $D^b_p(G)$ theories, where for a given arbitrary group $G$ and a parameter $b$, each theory in the same family has the same number of mass parameters, same number of marginal deformations, same $1$-form symmetry, and same $2$-group structure. This technique is utilized to establish the presence or absence of the 2-group symmetries in several classes of $D^b_p(G)$ theories. In this regard, we find that the $D_p($USp$(2N))$ theories constitute a special class of Argyres-Douglas theories that have a 2-group symmetry.}

\allowdisplaybreaks[1]

\begin{document} 

\maketitle

\section{Introduction}

Symmetry plays a crucial r\^ole as an organizing principle in the analysis of quantum field theory. The concept of symmetry has been evolving in the past several years since the introduction of the higher-form symmetries \cite{Kapustin:2014gua, Gaiotto:2014kfa} that act on higher dimensional operators. One of the important ideas is that a 0-form global symmetry and a 1-form global symmetry can be both present in several manners, for example as they can coexist as a direct product, there may be a mixed 't Hooft anomaly between them, or they can combine to form a non-trivial extension, known as a 2-group symmetry \cite{Sharpe:2015mja, Tachikawa:2017gyf, Benini:2018reh, Hsin:2020nts, Cordova:2020tij} (see also \cite{Argyres:2022mnu,Cordova:2022ruw} for a recent review).

This paper explores the presence of the 2-group symmetry in a broad family of 4d $\CN=2$ superconformal field theories (SCFTs), known as the Argyres-Douglas (AD) theories \cite{Argyres:1995jj,Argyres:1995xn,Eguchi:1996ds,Eguchi:1996vu}, whose feature is the presence of Coulomb branch (CB) operators with fractional conformal dimensions.\footnote{See also \cite{Gaiotto:2010jf,Giacomelli:2012ea} and \cite{Akhond:2021xio} for recent review on 4d SCFTs.} It turns out that there is a class of such theories known as the $D_p(\USp(2N))$ theories \cite{Wang:2018gvb,Carta:2021whq}, which have not received much attention in the literature, that possess a 2-group symmetry for an infinite family of values of $p$ and $N$.  Before going into further details, let us give a brief introduction into general AD theories.

\subsubsection*{$D^b_p(G)$ and $G^b[p]$ theories}
A class of theories that we  focus on is the $D^b_p(G)$ theories \cite{Cecotti:2012jx, Cecotti:2013lda, Wang:2018gvb}. If we consider the case of $b= h(G)$, the Coxeter number of $G$, then it is conventional to drop the superscript $b$ and write the corresponding theory as $D_p(G)$.  A $D^b_p(G)$ theory  can be realized in class $\mathcal{S}$ \cite{Gaiotto:2009we,Gaiotto:2009hg,Xie:2012hs,Wang:2018gvb} as well as from geometric engineering of Type IIB string theory \cite{Shapere:1999xr,Cecotti:2010fi}.  The former involves compactification of the 6d $\mathcal{N}=(2,0)$ theory on a sphere with a regular full puncture and an irregular puncture, where the compactification may involve an outer-automorphism twists when $G$ is non simply-laced.  The latter involves compactification of Type IIB string theory on a non-compact Calabi-Yau (CY) $3$-fold realized as the zero-locus of a single hypersurface singularity in $\mathbb{C}^3\times \mathbb{C}^\ast$. Given a $D^b_{p+b}(G)$ theory, the regular full puncture may be fully closed, and the resulting theory is conventionally denoted by $G^{b}[p]$, which can be realized from Type IIB string theory compactified on a hypersurface singularity in $\BC^4$.  In particular, if $G=A_n$ or $D_n$ or $E_n$, then $G^{b}[p]$ is identified with the well-known $(A_{p-1}, G)$ theory, written in the notation of \cite{Cecotti:2010fi}. More information about the $G^{b}[p]$ theories, with $G$ non simply-laced, can be found in \cite{Wang:2015mra, Carta:2021whq, Bhardwaj:2021mzl}.\footnote{In \cite{Bhardwaj:2021mzl} the $C^{2N}[p]$ theory is referred to as $\{7,1\}(2, p -2N +1,N,2)$.  This is described by the isolated hypersurface singularity $F(u,x,y,z) = u^2+x^{p-2N+1}+ xy^N+yz^2=0$ in $\BC^4$ with differential $\Omega = \frac{du dx dy dz}{dF}$. We will use the notation $C^{2N}[p]$ in this paper for compactness.} 

\subsubsection*{1-form symmetries}
One of the crucial ingredients that is needed in this paper is the information about the 1-form symmetries.  For the $(G,G')$ theories, where both $G$ and $G'$ are simply-laced, these have been worked out explicitly in \cite{Closset:2020scj, DelZotto:2020esg, Hosseini:2021ged, Closset:2021lwy, Buican:2021xhs} by means of the defect groups in Type IIB compactification on an isolated hypersurface singularity. Since this method can be generalized to general hypersurface singularities, one can compute the 1-form symmetries for general $G^b[p]$ theories, including non simply-laced $G$. From the class $\CS$ perspective, such a 1-form symmetry can also be realized as the defect group associated with the line defects {\it trapped} at the irregular puncture of the class $\CS$ realization\footnote{We remark that the 1-form symmetries of general class $\CS$ theories with regular punctures have been worked out in \cite{Bhardwaj:2021pfz}.} of the corresponding $G^b[p]$ theory \cite{Bhardwaj:2021mzl}. 

Given a $G^b[p]$ theory, there are two known ways to deduce the 1-form symmetries of the $D^b_p(G)$ and related theories. The first way is to exploit the following results: The 1-form symmetries are invariant under the Maruyoshi-Song \cite{Maruyoshi:2016tqk, Maruyoshi:2016aim, Agarwal:2016pjo} (see also \cite{Benvenuti:2017lle, Benvenuti:2017kud, Benvenuti:2017bpg}) flow \cite{Hosseini:2021ged} and are invariant under the Higgs branch (HB) flow \cite{Carta:2022spy}. For the former, the $D^b_p(G)$ theory flows to the $G^b[p]$ theory, whereas for the latter the $D^b_{p+b}(G)$ theory flows to the $G^b[p]$ theory upon closing the full puncture; see also \cite{Giacomelli:2017ckh}. We thus conclude that the $D^b_p(G)$, $D^b_{p+b}(G)$ and $G^b[p]$ theories all have the same 1-form symmetry. The other way is to use the powerful method of \cite{Bhardwaj:2021mzl}, which is elucidated in Section \ref{sec:bootstrap} of this paper: The $D^b_{c \, \GCD(p, b)}(G)$ theory has the same 1-form symmetry as the $G^b[p]$ and $D^b_p(G)$ theories, for any $c \geq 1$ such that $c$ is coprime to $\frac{b}{\GCD(p,b)}$.  Since this method involves generating an infinite family of theories from an elementary theory $D^b_{\GCD(p, b)}(G)$, we dub it ``bootstrapping''.  We point out in Section \ref{sec:bootstrap} that each theory in the same family not only has the same 1-form family, they also have the same number of mass parameters, same number of marginal deformations, same global form of the flavor symmetry associated with $G$, and same 2-group symmetry.

\subsubsection*{A global form of the flavor symmetry}

Another crucial ingredient is the global form of the flavor symmetry of a given $D^b_p(G)$ theory.  This theory has a global symmetry algebra $\mathfrak{g} \times \mathfrak{u}(1)^{m}$, where $\mathfrak{g}$ is the Lie algebra of the non-abelian group $G$, and $m$ is the extra number of mass parameters excluding the $G$ Casimirs.  Note that this symmetry may get enhanced further to a larger Lie algebra $\hat{\mathfrak{g}}$ that contains $\mathfrak{g} \times \mathfrak{u}(1)^{m}$ as a subalgebra, \eg $D_2(\SO(8))$ is known to be identical to the rank-1 $E_6$ theory \cite{Cecotti:2013lda}, whose flavor symmetry algebra is $\mathfrak{e}_6$.  In this paper, we are interested in determining the global form of flavor symmetry associated with $\mathfrak{g}$.  Note that for theories of class $\CS$ associated with a sphere with regular punctures, this was computed in \cite{Bhardwaj:2021ojs} (see also \cite{Distler:2020tub, Apruzzi:2021vcu, Distler:2022yse} for related discussions).  We propose that such a global form can be determined by the generators of the Higgs branch of the theory in question.\footnote{We remark that there are more rigorous ways to analyze the global form of the flavor symmetry, for example, using the superconformal index or its Schur limit. Although these quantities are available for some Argyres-Douglas theories \cite{Buican:2015tda, Buican:2017uka}, it is not available for general $D_p(\USp(2N))$ theories studied in our article. Of course, this is an interesting problem that is worth a future investigation.} In particular, if $\tilde{G}$ is the universal cover of $G$, $\mathbf{C}$ is the center of $\tilde{G}$, and $\CZ$ is a subgroup of $\mathbf{C}$ such that $\CZ$ acts trivially on the Higgs branch generators, it follows that the global form of the flavor symmetry associated with $G$ is given by $\tilde{G}/\CZ$.  As an example, for the rank-1 $E_6$ theory, the Higgs branch is the reduced one $E_6$ instanton moduli space, whose only generators are in the adjoint representation of $\mathfrak{e}_6$.  The global form of the flavor symmetry (not taking into account the $R$-symmetry) of this theory is, therefore, $E_6/\BZ_3$, since there is no generator transforming under the $\BZ_3$ center of $E_6$; this conclusion is in agreement with \cite[(4.41)]{Bhardwaj:2021ojs}.\footnote{Note however that, as pointed out in \cite[Section 3.3]{Heckman:2022suy}, there is a mixing between the center of the $E_6$ flavor symmetry and the $\U(1)_R$ symmetry of this rank-1 $E_6$ SCFT, and the correct form of the global symmetry after taking into account the $R$-symmetry is $[E_6 \times \U(1)_R]/\BZ_3$. This can be seen also from the fact that there are BPS states in the fundamental representation $\mathbf{27}$ of $E_6$ \cite{Distler:2019eky} that do not contribute either to the superconformal index \cite{Gadde:2010te} nor the Higgs branch of the theory. We thank Craig Lawrie for pointing this out to us.}  We can now determine the global form of the flavor symmetry associated with $\so(8)$ in $D_2(\SO(8))$.  What we need to do is to compute the branching rule of the adjoint representation of $\mathfrak{e}_6$ into representations of $\so(8) \times \mathfrak{u}(1)^2$. The result are the adjoint, vector, spinor and cospinor representations of $\so(8)$, and so we conclude that the global form of the flavor symmetry associated with $\so(8)$ is $\Spin(8)$.  In many cases, it is simple to work out the generators of the Higgs branch of a given 4d theory from the Coulomb branch of the corresponding 3d mirror theory, since for $D^b_p(G)$ theories with $G=\SU(N)$ and $\SO(2N)$, the mirror theories admit 3d $\CN=4$ Lagrangian description in terms of quiver gauge theories; see \cite{Giacomelli:2020ryy, Carta:2021whq, Carta:2021dyx}.  In this paper, we present also the 3d mirror theories for general $D_p(\USp(2N))$ theories.  These turn out to be very useful for the study of the Higgs branch of the latter.

\subsubsection*{2-group symmetries}
Another protagonist of this paper is the 2-group symmetry that involves a non-trivial extension between a discrete 1-form symmetry and a continuous 0-form symmetry. This type of 2-group symmetry has been studied in a wide range of theories, see \eg \cite{Hsin:2020nts, Bhardwaj:2021wif, Lee:2021crt, Apruzzi:2021mlh,DelZotto:2022fnw, DelZotto:2022joo,Cvetic:2022imb, Bhardwaj:2022scy}. One of the important theories is the 4d $\CN=2$ $\Spin(4n+2)$ gauge theory with $N_f$ hypermultiplets of vector hypermultiplets. Although the center of the $\Spin(4n+2)$ group is $\BZ_4$, this theory has an electric $\BZ_2$ 1-form symmetry, which is the group of the gauge Wilson lines after taking into account of the screening. The global form of the 0-form flavor symmetry is $\USp(2N_f)/\BZ_2$, where the $\BZ_2$ quotient arises from the fact that  there is no Higgs branch generator charged under the center of the $\USp(2N_f)$ flavor symmetry. It was shown in \cite{Lee:2021crt} (see also \cite{Apruzzi:2021mlh}) that this theory has a 2-group symmetry that is a non-trivial extension between the electric $\BZ_2$ 1-form symmetry and the $\USp(2N_f)/\BZ_2$ 0-form flavor symmetry.  This is characterized by the following short exact sequence that does not split:
\bes{ \label{exactseq1}
0 \rightarrow \Gamma^{(1)} \rightarrow \CE \rightarrow \CZ \rightarrow 0
}
where $\Gamma^{(1)}=\BZ_2$ corresponds to the 1-form symmetry;  $\CZ= \BZ_2$ corresponds to the quotient in $\USp(2N_f)/\BZ_2$; and  $\CE =\BZ_4$ in the middle of the exact sequence corresponds to the maximally trivially acting group on the matter fields, which can be realized as follows. The vector representation transforms as $-1$ under the generator $e^{i \pi/2}$ of the $\BZ_4$ center of the gauge group $\Spin(4n+2)$.  The hypermultiplets also transform as $-1$ under the center $e^{i \pi}$ of the $\BZ_2$ center of the flavor symmetry $\USp(2N_f)$.  Therefore, the diagonal combination $\omega \equiv (e^{i \pi/2}, e^{i \pi})$ leaves the matter fields invariant. The element $\omega$ can indeed be identified with the generator of $\CE = \BZ_4$, which is a non-trivial extension between $\Gamma^{(1)}=\BZ_2$ and $\CZ= \BZ_2$.  If $w_2$ is the 2nd Stiefel-Whitney class obstructing the $\USp(2N_f)/\BZ_2$ bundles to the $\USp(2N_f)$ bundles and $B_2$ is the background field associated with the 1-form symmetry, then $\delta B_2 = \mathrm{Bock}(w_2)$, where $\mathrm{Bock}$ is the Bockstein homomorphism for the exact sequence \eref{exactseq1}.

A generalization of this result to a wider class of 4d $\CN=2$ SCFTs is to consider the $D_p(\USp(2N)) \equiv D^{2N}_p(\USp(2N))$ theories, which is engineered by the hypersurface singularity in $\mathbb{C}^3\times \mathbb{C}^*$ and holomorphic differential %{\color{red} TO BE FIXED} 
(see \eg \cite[(2.33)]{Carta:2021whq})
\bes{
%F(u,x,y,z)=u^2+ x^{p+1-2N}+ xy^{N-1} + yz^2=0 ~, \qquad \Omega =\frac{du dx dy dz}{dF}~. %AM: This was the one after closing the puncture.
F(u,x,y,z)=u^2+ x^{N}+ xy^{2} + z^p=0 ~, \qquad \Omega =\frac{du dx dy dz}{zdF}~,
}
where $z$ is the $\mathbb{C}^*$ variable. These theories can also be realized as twisted-compactification of the 6d $\CN=(2,0)$ of the $D_{N+1}$-type on a sphere with a twisted irregular puncture and a twisted regular full puncture, where the latter is labeled by the $C$-partition $[1^{2N}]$.  In fact, the $D_2(\USp(8n))$ theory corresponds to a Lagrangian theory described by the $\so(4n+2)$ gauge theory with $4n$ hypermultiplets in the vector representation.  As described in \cite{Bhardwaj:2021wif}, we can choose the group of the genuine line defects (modulo screening of the matter fields), also known as the polarization, in such a way that the class $\CS$ theory in question becomes an absolute theory. There is a choice that corresponds to fixing the gauge group to be $\Spin(4n+2)$; in which case the theory has a 2-group symmetry for the reasons described above.\footnote{On the other hand, if the gauge group is chosen to be $\SO(4n+2)$, this corresponds to gauging the 1-form symmetry of the $\Spin(4n+2)$ gauge theory; in which case there is a mixed anomaly \cite{Lee:2021crt} between the new 1-form magnetic symmetry of the $\SO(4n+2)$ gauge theory and the flavor symmetry $\USp(2N_f)/\BZ_2$, with $N_f = 4n$.}  Exploiting the result of \cite{Bhardwaj:2021mzl},  we propose that this can then be generalized to a wider class of theories.  Specifically, we propose that each of the $D_{\mathfrak{p}c}(\USp(2M\mathfrak{p}))$ theories, with $\mathfrak{p}$ and $M$ even and $c$ coprime to $2M$, has a 1-form symmetry $\BZ_2^{\mathfrak{p}/2}$, and if this is chosen to be the electric 1-form symmetry then it forms a 2-group structure with the 0-form flavor symmetry $\USp(2M\mathfrak{p})/\BZ_2$.  We also explore certain dynamical consequences of this proposal.

\subsubsection*{Organization of the paper}

The paper is organized as follows. In Section \ref{sec:bootstrap} we discuss the ``bootstrap'' method that generates a family of theories with the same number of mass parameters, same number of marginal deformations, same 1-form symmetry and same 2-group structure. In Section \ref{sec:DpUSp12sym} we discuss the $D_p(\USp(2N))$ theories, along with their 1-form symmetries, 2-group structures and 3d mirror theories. We organize the discussion according to the number of mass parameters of the theory.  In Appendix \ref{sec:abseDpG}, we consider other $D^b_p(G)$ theories, including the $D_p^{N+1}(\USp(N))$, $D_p^b(\USp'(2N))$ and $D_p^b(\SO(2N+1))$ theories.  We found that these theories do not have a 2-group symmetry.   {\it This indeed makes $D_p(\USp(2N))$ a special class of Argyres-Douglas theories when it comes to the presence of the 2-group structure.}  We end this paper by discussing how the Higgs branch structure of the $D_p(\SU(N))$ theories changes under bootstrap in Appendix \ref{sec:HBbootstrap}.

\subsubsection*{Notation and convention}
We adopt the following notation and convention.
\bi
\item Upon discussing the number of mass parameters of the $D^b_p(G)$ theories, we exclude the Casimirs of $G$ for convenience. In particular, it was pointed out in \cite[Appendix B.6]{Giacomelli:2017ckh} that the number of mass parameters of the $J^b[p]$ theory is equal to that of the $D^b_p(G)$ theory, excluding the Casimirs of $G$.
\item Unless stated explicitly otherwise, in the quiver diagram we denote the orthosymplectic gauge algebras by $B_n = \so(2n+1)$, $C_n = \usp(2n)$ and $D_n = \so(2n)$, and in these cases the gauge groups are taken to be their universal covers, namely $\Spin(2n+1)$, $\USp(2n)$ and $\Spin(2n)$, respectively.
\ei

\section{Bootstrapping \texorpdfstring{$D^b_p(G)$}{} theories} \label{sec:bootstrap}
In this section, we present a method of obtaining an infinite family of AD theories starting from an arbitrary $D^b_p(G)$ theory, where the theories in the same family share a number of interesting properties. The main idea was introduced in \cite{Bhardwaj:2021mzl}. We elucidate it as follows. 

We consider a $D^b_p(G)$ theory, such that $(G,b)$ takes one of the following values: 
\bes{ \label{Gblist}
\scalebox{0.96}{\renewcommand{\arraystretch}{1.5}
\begin{tabular}{llll}
$\left(\SU(N), \substack{N \\ N-1} \right)$ \, & $\left(\SO(2N+1), \substack{2N \\ 2N-1^*} \right)$ \, & $\left(\SO(2N), \substack{2N-2 \\ N}\right)$ \, & $\left(\USp(2N), \substack{2N \\ N+1^*}\right)$ \\
$\left(E_6, \substack{12 \\ 9 \\ 8} \right)$ \, & $\left(E_7, \substack{18 \\ 14} \right)$ \, & $\left(E_8, \substack{30 \\ 24\\ 20} \right)$ & \\
$\left(F_4, \substack{12 \\ 18 \\ 8} \right)$ \, & $\left(G_2, \substack{6 \\ 12} \right)$ & $\left(\USp'(2N), \substack{2N \\2N+1^*} \right)$ \, &
\end{tabular}
}
}
For each $G$, there are many values of $b$ listed in the column next to it.  The top line in each column corresponds to $b=h(G)$, the Coxeter number of $G$; in which case, we drop the superscript $b$ and write $D_p(G) \equiv D^{h(G)}_p(G)$.\footnote{Note, however, that $D^{2N+1}_p(\USp'(2N))$ was denoted by $D_p(\USp'(2N))$ in \cite{Carta:2021whq}. In order to avoid the confusion,  in this paper we explicitly write it as $D^{2N+1}_p(\USp'(2N))$.}  Note that for $(G,b)= (\USp(2N), N+1)$, $(\USp'(2N), 2N+1)$ and $(SO(2N+1), 2N-1)$, $p$ is {\it half-odd-integral} (see \cite[(2.15), (2.23)]{Carta:2021whq} and also Appendices \ref{sec:abseDpUSp} and \ref{sec:twistedA}) and these are emphasized by ${}^*$ in the above list; in the other cases $p$ is integral.  We refer the reader to \cite[Tables 2 and 3]{Bhardwaj:2021mzl} for more information.\footnote{Note that in \cite[Table 3]{Bhardwaj:2021mzl}, the rows with $b_t = 4N+2$ (twisted $A_{2N}$), $b_t = 4N-2$ (twisted $A_{2N-1}$), and $b_t = 2N+2$ (twisted $D_{N+1}$) correspond to $D^{2N+1}_p(\USp'(2N))$, $D^{2N-1}_p(\SO(2N+1))$, and $D^{N+1}_p(\USp(2N))$, each with $p$ half-odd-integral, in our notation.}

The spectral equation for the Hitchin field on the Riemann sphere implies that the monodromy is fully encoded in the quantity
\bes{
q = \frac{b}{\GCD(p, b)}~.
} 
when $p$ is an integer, whereas
\bes{
q = \frac{2b}{\GCD(2p, 2b)}~.
}
whenever $p$ is half-odd-integral, namely for $(G,b)= (\USp(2N), N+1)$, $(\SO(2N+1), 2N-1)$ and $(\USp'(2N), 2N+1)$. We examine these cases in Appendices \ref{sec:abseDpUSp} and \ref{sec:twistedA}. Notice that for integer $p$ the two definitions of $q$ agree.

Observe that $q$ is invariant under replacing $p$ by $\frac{\GCD(2p,2b)}{2}$.  In fact, $q$ is invariant under replacing $p$ by $c\, \frac{\GCD(2p,2b)}{2}$ for any $c\geq 1$ such that $c$ is coprime to $q$.

Utilizing this observation, we start with the $D^b_{\GCD(2p, 2b)/2}(G)$ theory, and then construct the following {\it family of theories} 
\bes{ \label{family}
 \left\{ D^b_{c \, \GCD(2p, 2b)/2}(G) \left| \text{$c$ is coprime to $q \equiv \frac{2b}{\GCD(2p, 2b)}$} \right. \right\}\fstop
 } 
As described above, the theories belonging to the same family have the same parameter $q$ controlling the monodromy of the Hitchin field. We will refer to this procedure of constructing the family of theories as ``bootstrap'', where the ``smallest'' theory in this family is that with $c=1$, namely $D^b_{\GCD(2p, 2b)/2}(G)$.

From now on we will focus on the case $G$ classical. The advantage is that whenever $q$ is even, the smallest $D_{\GCD(2p, 2b)/2}(G)$ theory in the family admits a 4d $\CN=2$ Lagrangian description \cite{Bhardwaj:2021mzl}, whereas this is not the case for $q$ odd (unless $G=\SU(N)$). For $c>1$, the $D_{c \, \GCD(2p, 2b)/2}(G)$ theory does not necessarily have a $\CN=2$ Lagrangian description.

It was pointed out in \cite{Bhardwaj:2021mzl} that the theories in the same family has the same 1-form symmetry.  In this paper, we also observe that each theory in a given family has
\ben
    \item the same number of mass parameters, \label{obsmass}
    \item the same number of marginal deformations, \label{obsconf}
    \item the same global form of the flavor symmetry associated with $G$, and \label{obsglobalform}
    \item the same 2-group structure.\label{obs2group}
\een
For $G=\SU(N)$, $\SO(2N)$ and $E_{6,7,8}$ with any allowed values of $b$ listed in \eref{Gblist}, Observation \ref{obsmass} can be seen explicitly from Table 1 in \cite[Appendix B]{Giacomelli:2017ckh}.  In this paper, we find that this is also the case for $G=\USp(2N)$. 

For $G=\SU(N)$ and $\SO(2N)$ with any allowed values of $b$ listed in \eref{Gblist}, Observation \ref{obsconf} follows from \cite{Giacomelli:2020ryy, Carta:2021whq, Carta:2022spy}. As an example, the number $x$ of exactly marginal deformations for the $D_p(\SO(2N))$ theory is explicitly given by \cite[(3.3)]{Carta:2021whq}: If $(2N-2)$ divides $Np$ then $x=\GCD(N-1,p)$, and if $(2N-2)$ does not divide $Np$, then $x=\GCD(N-1,p)-1$. Observe that for a given family \eref{family}, each theory belongs to the same (one or the other) case, and that $\GCD(N-1, c \, \GCD(p, 2N-2)) = \GCD(N-1, \GCD(p, 2N-2))$, independently of $c$; therefore, each theory in the family has equal $x$. In this paper, we find that Observation \ref{obsconf} also holds for $G=\USp(2N)$.

We now turn to Observation \ref{obsglobalform}.  Let us consider an example of the $D_p(\SU(N))$ theories with zero mass parameter, excluding the Casimirs of $\SU(N)$. A necessary and sufficient condition for this is $\GCD(p,N)=1$ \cite{Giacomelli:2017ckh}, and so any theory with this number of mass parameter can be bootstrapped from the $D_1(\SU(N))$ theory, which is a theory of free $N$ hypermultiplets. The global form of the flavor symmetry $\su(N)$ of each theory in this class is $\SU(N)/\BZ_N$, since the only generators of the Higgs branch are in the adjoint representation of $\SU(N)$ and so there is no operator charged under the $\BZ_N$ center. The latter can be seen from the 3d mirror theory, whose interacting part is the $T[\SU(N)]$ theory \cite{Giacomelli:2020ryy}; indeed, the only generators of the Coulomb branch are in the adjoint representation of $\SU(N)$. As an important remark, for some $D_p(G)$ theory with a sufficiently small $p$, its flavor symmetry $\hat{G}$ may be larger than $G$ and contain $G$ as a subgroup; in which case, upon decomposing the representations of $\hat{G}$ under which the Higgs branch generators transform into those of $G$, one can determine the global form of the flavor symmetry $G$ and we observe that the latter is the same for each theory in the same family. For example, the $D_2(\USp(2))$ theory is known to be identical to the $(A_1, D_4)$ theory, whose flavor symmetry algebra is $\su(3)$. The Higgs branch is the reduced moduli space of one $\su(3)$ instanton, where the only generators of the Higgs branch are the moment map in the adjoint representation $[1,1]$ of $\su(3)$. The global form of the $\su(3)$ flavor symmetry algebra is, therefore, $\SU(3)/\BZ_3$. Upon decomposing $[1,1]$ of $\su(3)$ into $\usp(2) \times \mathfrak{u}(1)$, we obtain $[2]_0+[1]_3+[1]_{-3}+[0]_0$, and so the global form of the $\usp(2)$ global symmetry algebra of the $D_2(\USp(2))$ theory is actually $\USp(2)$ and not $\USp(2)/\BZ_2$, since the fundamental representation $[1]$ transforms non-trivially under the $\BZ_2$ center.  To see that the global form of the flavor symmetry of $D_{2c}(\USp(2)) \cong (A_1, D_{4c})$ for any integer $c \geq 1$ is indeed $\USp(2)$ and not $\USp(2)/\BZ_2$, we exploit the Coulomb branch of the 3d mirror theory of the $(A_1, D_{4c})$ theory, which was studied in \cite[(4.17)]{Carta:2021whq} and is depicted in \eref{mirrDpUSp2b} of this paper.\footnote{The Coulomb branch Hilbert series of \eref{mirrDpUSp2b} is 
\bes{ \label{HSA3Dp}
H(t; x, u)\sum_{m_1 \in \BZ} \sum_{m_2 \in \BZ} t^{|m_1|+|m_1-m_2|+(p-1)|m_2|} (1-t^2)^{-2} x^{2m_1} (u^3 x^{-1})^{m_2}~,
}
where $x$ is the $\USp(2)$ fugacity and $u$ is the $\U(1)$ fugacity.  For $p=2$, the order $t^2$ of this Hilbert series receives the contribution of the moment map in the adjoint representation of $\SU(3)$ written in terms of $\USp(2) \times \U(1)$: $(u^3+ u^{-3})(x+x^{-1})+(x^2+1+x^{-2})+1$, as it should be.
}  Using the Coulomb branch Hilbert series or the index, it can be shown that there always exist Coulomb branch generators in the representations $[1]_{\pm 3}$ of $\usp(2) \times \mathfrak{u}(1)$ symmetry.\footnote{These arise from the magnetic fluxes $(m_1, m_2) = \pm(0,1)$ and $\pm(1,1)$ for the left and right $\U(1)$ gauge groups, using in the notation of \eref{HSA3Dp}.} This means that, for any $c\geq 1$, the global form of the $\usp(2)$ symmetry algebra for the $D_{2c}(\USp(2))$ theory is $\USp(2)$ and not $\USp(2)/\BZ_2$.

Let us now discuss the observation regarding the 2-group structure.  Since the 1-form symmetry and the global form of the flavor symmetry algebra are the same for each theory in the same family, if the former is trivial or the latter is simply-connected for a theory in the class, then we can rule out the existence of a 2-group structure for the whole family.  However, if the 1-form symmetry is non-trivial and there is a non-trivial quotient $\CZ$ in the global form of the flavor symmetry, we need to determine whether is a non-trivial extension between the 1-form symmetry and the obstruction class controlling whether the $G/\CZ$ bundle lifts to the $G$ bundle. 
If a theory in a given family \eref{family} admits the 4d $\CN=2$ Lagrangian description, one can be determined whether this theory admits a 2-group structure using the standard method, such as in \cite{Tachikawa:2017gyf, Hsin:2020nts, Apruzzi:2021mlh}.  To further conclude that the other theories in the family have the same 2-group structure, we utilize the result of \cite{Bhardwaj:2021mzl}, whose idea is as follows.  For a general theory of class $\CS$ associated with a (twisted) sphere with one (twisted) irregular and one (twisted) full puncture, the electric/magnetic 1-form symmetry was identified with the trapped electric/magnetic part $\CH_\CP^T$ of the defect group associated with the irregular puncture $\CP$, and the extension (\ie the maximally trivially acting group) was identified with the electric/magnetic part $\CH_\CP$ of the defect group of the irregular puncture $\CP$. Reference \cite{Bhardwaj:2021mzl} provided information about the following short exact sequence
\bes{ \label{exactseq2}
0~ \rightarrow~ \CH_\CP^T ~ \rightarrow~ \CH_\CP ~ \rightarrow~ \cZ ~\rightarrow 0~,
}
which is believed to contain the same information as \eref{exactseq1}. We shall provide a justification for this in Section \ref{sec:justification}. Due to the fact that each theory in a given family has the same monodromy data of the Higgs field, \eref{exactseq2} holds for every theory in the same family. In particular, if the above short exact sequence does not split, we conclude that every theory in the family has a non-trivial 2-group structure.  On the other hand, if a Lagrangian theory in the family does not have a 2-group symmetry, the above exact sequence splits for every theory in the family.  We test these statements by studying the dynamical consequences in Section \ref{sec:dynconsequences}.

\subsection{Bootstrapping 3d mirror theories of \texorpdfstring{$D_p(G)$}{}}
\label{sec:bootstrapmirror}
Let us consider how 3d mirror theories of $D_p(G)$ transform under bootstrapping.

\subsubsection{The case of \texorpdfstring{$G=\SU(N)$}{}}
Let us consider the $D_p(\SU(N))$ theory with $b=h(\SU(N))=N$. For simplicity, we focus on those cases with $p\geq b$. Let us call
\begin{equation}
    \mu=\GCD(p,N)=\GCD(p,b)\coma N= q\mu \coma p = c \mu\coma
\end{equation} 
where $c$ and $q$ are indeed coprime. This theory can therefore be obtained by bootstrapping the $D_\mu(\SU(q \mu))$ theory. As discussed in \cite[Section 4]{Giacomelli:2020ryy}, the $3$d mirror of the $D_p(\SU(N)) = D_{\mu c}(\SU(q \mu))$ theory is given by a complete graph with $\mu$ $\U(1)$ vertices with equal edge multiplicity $\fm \equiv q(c-q)$, where each $\U(1)$ is connected to the $\U(N-1)$ node of a $\U(N-1)-\U(N-2)-\cdots -\U(2)-\U(1)$ tail with edge multiplicity $q$. Moreover, there are a number of free hypermultiplets equal to $H_{\text{free}} \equiv \frac{1}{2}\mu(q-1)(c-q-1)$. Decoupling the tail, we obtain the $3$d mirror theory for $(A_{\mu(c-q)-1},A_{\mu q-1})$, which is a complete graph of $\mu$ $\U(1)$ with multiplicities $q(c-q)$ and $H_{\text{free}}$ free multiplets. We emphasize that the parameter $c$ enters in two places: the edge multiplicity $\fm$ of the complete graph, and the number $H_{\text{free}}$ of free hypermultiplets.  We see that, under bootstrapping, the changes in $\fm$ and $H_{\text{free}}$ are
\bes{
\delta \fm = q(c-q) - q(1-q)= \frac{q}{\mu} \delta p~, \qquad 
\delta H_{\text{free}} = \frac{1}{2} (q-1) \delta p~.
\label{eq:DpSUhyper}
}
where
\bes{
\delta p = \mu c -\mu = \mu (c-1)~.
}

It is instructive to compare these changes with the Maruyoshi-Song flow \cite{Maruyoshi:2016tqk, Maruyoshi:2016aim, Agarwal:2016pjo, Giacomelli:2017ckh} discussed in \cite[Section 4.3]{Giacomelli:2020ryy}, where in the latter we have
\bes{
\delta p = b= q \mu~, \quad \delta \fm = q^2 =  \frac{q}{\mu} \delta p~, \quad \delta H_{\text{free}} = \frac{1}{2} \mu q(q-1) = \frac{1}{2}(q-1) \delta p~. 
}
It is therefore clear that the shifts of the parameters under the Maruyoshi-Song flow is a {\it special case} of the bootstrap with $c=q+1$.

\subsubsection{The case of \texorpdfstring{$G=\SO(2N)$}{}}
Let us now consider the $D_p(\SO(2N))$ theory with $b= h(\SO(2N)) = 2N-2$. There are three cases to be considered according to the number of mass parameter of the theory.\footnote{We stress again that in the following we exclude the Casimirs of $\SO(2N)$ in the counting of the mass parameters.}  

\subsubsection*{Zero mass parameter}
A necessary and sufficient condition for a $D_p(\SO(2N))$ theory to have zero mass parameter is that $\GCD(p, b) = \GCD(p, 2N-2)$ is odd. For convenience, we define
\bes{ \GCD(p, 2N-2) = 2\mu-1~, \qquad q=\frac{2N-2}{2\mu-1}~. }
As described in \cite[Section 5]{Carta:2021whq}, the mirror theory is described by the $T[\SO(2N)]$ theory together with 
\bes{
H_{\text{free}} = \frac{1}{2} N(p-2N+1)
}
free hypermultiplets. Similarly to the case of $D_p(\SU(N))$, we see that, bootstrapping from $D_{2\mu-1}(\SO(2N))$ to $D_{(2\mu-1)c}(\SO(2N))$, such that $\GCD(c, q)=1$, we see that $H_{\text{free}}$ gets shifted by 
\bes{
\delta H_{\text{free}} = \frac{1}{2} N \delta p~, \quad \text{with}~~\delta p = (c-1)(2\mu-1)~.
}

Under the Maruyoshi-Song flow \cite{Maruyoshi:2016tqk, Maruyoshi:2016aim, Agarwal:2016pjo, Giacomelli:2017ckh} discussed in \cite[Section 5]{Carta:2021whq}, we have
\bes{
\delta H_{\text{free}} = N(N-1) = \frac{1}{2} N \delta p ~, \quad \text{with}~~\delta p = b =2N-2~.
}
We see again that the shifts of the parameters under the Maruyoshi-Song flow is a {\it special case} of the bootstrap with $c=q+1$. 

\subsubsection*{One mass parameter}
A necessary and sufficient condition for a $D_p(\SO(2N))$ theory to have one mass parameter is that \bes{
\GCD(p, b) = \GCD(p, 2N-2) \equiv 2 \mu 
}
is even, and 
\bes{
q \equiv \frac{2N-2}{\GCD(p, 2N-2)} =\frac{N-1}{\mu}
}
is even. As pointed out in \cite[(6.7)]{Carta:2021whq}, the 3d mirror theory of such a $D_p(\SO(2N))$ theory is 
\bes{
H_{\text{free}}= \frac{1}{2}\left[ N(p-2N+5)-2p-3 \right]
}
together with an interacting part described by a quiver consisting of the tail $D_1-C_1-\cdots-D_{N-1}-C_{N-1}$ such that the $C_{N-1}$ node has $N-1$ flavors of fundamental hypermultiplets and is connected with an edge with multiplicity $1$ to a $D_1$ node with $x \equiv \frac{1}{2}[p-(2N-2)]$ hypermultiplets of charge one.  We see that bootstrapping from the $D_{2\mu} (\SO(2N))$ theory to $D_{2\mu c} (\SO(2N))$ such that $\GCD(c, q)=1$ shifts the parameters as follows:
\bes{
\delta p = 2\mu(c-1)~, \quad \delta x =\frac{1}{2}\delta p ~, \quad  \delta H_{\text{free}} = \frac{1}{2} (N-2) \delta p~.
}
Under the Maruyoshi-Song flow described in \cite[Section 6.1.1]{Carta:2021whq}, $\delta p =b =2N-2 = 2 \mu q$ and the parameters $x$ and $H_{\text{free}}$ get shifted in the same way.  Therefore, it is a special case of the bootstrap with $c=q+1$.

\subsubsection*{More than one mass parameter}
Let us now consider the $D_p(\SO(2N))$ theories with $p \geq b=2N-2$ and $\mu+1$ mass parameters such that $\mu \geq 1$.  These correspond to the case where
\bes{
\GCD(p,b)= \GCD(p,2N-2)= 2\mu 
}
is even, and
\bes{
q= \frac{2N-2}{\GCD(p,2N-2)} = \frac{N-1}{\mu}
}
is odd.  As explained in \cite[Section 6.1.2]{Carta:2021whq}, the mirror of such a theory consists of
\bes{
H_{\text{free}} = \frac{1}{4\mu}(N-1-\mu)[p-(2N-2)-2\mu] = \frac{1}{4}(q-1)(p-b-2\mu)
}
free hypermultiplets and the interacting part described by a quiver that contains the tail $D_1-C_1-D_2-C_2-\cdots-C_{N-1}$ connecting with $\mu$ $D_1$ nodes of two types: $\mu$ nodes of Type A and $1$ node of Type B. Each of the $D_1$ nodes of Type A is connected to the $C_{N-1}$ node in the tail with equal edge multiplicity $q$ and has $F=\frac{1}{4 \mu}(q-1)(p-b)$ flavors of hypermultiplets of charge 1. On the other hand, the $D_1$ node of Type B is connected to the $C_{N-1}$ node in the tail with multiplicity $1$. The nodes of Type A are connected to each other and form a complete graph with edge multiplicity $M = \frac{1}{2}q(p-b)$. The node of Type B is connected to each node of Type A by equal edge multiplicity $\fm=\frac{1}{2 \mu}(p-b)$.

Bootstrapping from the $D_{2\mu}(\SO(2N))$ theory to the $D_{2\mu c}(\SO(2N))$ theory, such that $\GCD(c, q)=1$, we see that the parameters are shifted by
\bes{
\delta H_{\text{free}} = \frac{1}{4}(q-1)\delta p~, \quad \delta F = \frac{1}{4\mu} (q-1) \delta p~, \quad \delta M = \frac{1}{2} q \delta p~, \quad \delta \fm = \frac{1}{2\mu}\delta p~,
}
where
\bes{
\delta p = 2\mu(c-1)~.
}
As before, the Maruyoshi-Song flow discussed in \cite[Section 6.1.2]{Carta:2021whq} corresponds to the special case with $c=q+1$.

\section{\texorpdfstring{$D_{p}(\USp(2N))$}{} theories: 1-form and 2-group symmetries}
\label{sec:DpUSp12sym}
In this section, we focus on the $D_{p}(\USp(2N))$ theories and their features, namely 1-form and 2-group symmetries. The main results in this section are as follows:
\bi
\item Every $D_{p}(\USp(2N))$ theory with zero mass parameter can be bootstrapped from a Lagrangian theory $D_{\fp}(\USp(2M \fp))$ with $M \geq 1$ and $\fp$ even.  In other words, every theory with zero mass parameter can be written as $D_{\fp c}(\USp(2M \fp))$ with $c$ coprime with $q \equiv 2M$.

If $M$ is even (\ie~ $q$ is divisible by $4$), it follows from the reasoning in \cite{Lee:2021crt, Bhardwaj:2021mzl, Apruzzi:2021mlh} that there is a choice of polarization such that the corresponding theory has a  {\it non-trivial 2-group symmetry}.\footnote{As an example, for $\mathfrak{p}=2$, the theory in question reduces to $D_2(\USp(4M))$, which is the $\so(2M+2)$ gauge theory with $2M$ hypermultiplets in the vector representation. If we choose the gauge group to be $\Spin(2M+2)$, then, as explained in \cite{Lee:2021crt}, this theory has a $2$-group structure for $M$ even. \label{foot:example}} 

On the other hand, if $M$ is odd (\ie~ $q$ is even but is not divisible by $4$), the corresponding theory does not have a 2-group structure.
\item For the theory with one mass parameters, those that can be bootstrapped from an interacting Lagrangian theory have a non-trivial 1-form symmetry, whereas those that cannot have a trivial 1-form symmetry.  All theories with two or more mass parameters have a trivial 1-form symmetry.
\item All theories with one and higher mass parameters have a trivial 2-group symmetry.    
\ei

\subsection{Zero mass parameter}
We find that the necessary and sufficient condition for a $D_p(\USp(2N))$ theory to have zero mass parameter is that
\bes{
p = \fp c~,\quad  N= M \fp~, \quad \fp~\text{even}~, \quad \GCD(c, 2M)=1~.
}
We will take $\fp$ to be even throughout this subsection.  Note that, when $c=1$, the $D_{\fp}(\USp(2M \fp))$ theory admits the following Lagrangian description \cite[(A.3)]{Carta:2021dyx}:
\bes{ \label{LagDpUSp2Mp}
[C_{M \fp}]-D_{M \fp-(M-1)}-C_{M \fp-2M}- \cdots -D_{3M+1}-C_{2M}-D_{M+1}
}
Note that, for $c >1$, $D_{\fp c}(\USp(2M\fp))$ does not admit a $\CN=2$ Lagrangian description.

\subsubsection{\texorpdfstring{$M$}{} even and the 2-group structure}

Let us take $M$ to be even and consider theory \eref{LagDpUSp2Mp}. In this case, all the $D$-type gauge groups in \eref{LagDpUSp2Mp} have odd rank.  For definiteness, we take each $D_{2m+1}$ gauge group to be $\Spin(4m+2)$.\footnote{We emphasize that this is a choice that gives rise to a non-trivial 2-group symmetry.  If one, on the other hand, takes all $D_{2m+1}$ gauge group to be $\SO(4m+2)$, then this is equivalent to gauging the $\Gamma^{(1)} = \BZ_2^{\fp/2}$ electric 1-form symmetry of the theory in question. As a result, we obtain the new $\BZ_2^{\fp/2}$ magnetic 1-form symmetry. There is a mixed anomaly between the latter and the 0-form flavor symmetry. As an example, we can consider the $D_2(\USp(4M))$ theory, as discussed in Footnote \ref{foot:example}: If we choose the gauge group corresponding to the $\so(2M+2)$ gauge algebra to be $\SO(2M+2)$, then there is a mixed anomaly between the $\BZ_2$ magnetic 1-form symmetry and the flavor symmetry $\USp(4M)/\BZ_2$ \cite{Lee:2021crt}.} Each spin gauge group has a $\BZ_4$ center and each symplectic gauge group has a $\BZ_2$ center, but they are screened to $\BZ_2$ and a trivial group by matter in the vector representation and fundamental representation, respectively. The electric $1$-form symmetry of theory \eref{LagDpUSp2Mp} is, therefore, $\Gamma^{(1)} = \BZ_2^{\fp/2}$, where $\fp/2$ is the number of the $D$-type gauge groups.
Since the gauge invariant operators of \eref{LagDpUSp2Mp} transform under adjoint representation of $C_{M \fp}$, the global form of the flavor symmetry group of this theory is $\USp(2M \fp)/\BZ_2$, where the $\BZ_2$ quotient comes from the fact that there are no operators charged under the $\BZ_2$ center of $\USp(2M \fp)$. 
According to \cite{Lee:2021crt, Apruzzi:2021mlh}, theory \eref{LagDpUSp2Mp} has a non-trivial 2-group structure, characterized by the following short exact sequence that does {\it not} split:
\bes{ \label{exactsequence}
0~ \rightarrow~ \Gamma^{(1)} ~ \rightarrow~ \mathcal{E} ~ \rightarrow~ \cZ ~\rightarrow 0\coma 
}
where the 1-form symmetry $\Gamma^{(1)}$, the maximally trivially acting group $\cE$, and the subgroup $\CZ$ of the center symmetry of $\USp(2M\fp)$ are given by
\bes{
\Gamma^{(1)}=\BZ^{\fp/2}_2 = \BZ^{(\fp-2)/2}_2 \times \BZ_2~, \quad \cE =\BZ^{(\fp-2)/2}_2 \times \BZ_4 ~, \quad \cZ = \BZ_2~.
}
The corresponding Postnikov class is given by $\mathrm{Bock}(w_2)$, where $w_2$ is the obstruction class for lifting the $\USp(2M \fp)/\BZ_2$ bundle to the $\USp(2M \fp)$ bundles, and $\mathrm{Bock}$ is the Bockstein homomorphism associated with \eref{exactsequence}.

Let us now consider the $D_{\fp c}(\USp(2M \fp))$ theory, with $\fp$ and $M$ even such that $\GCD(c, 2M)=1$. These theories can be realized as a theory of class $\CS$ of twisted $D$-type with a regular twisted puncture and an irregular twisted puncture $\CP$. Many of their properties were discussed in \cite[Section 9]{Bhardwaj:2021mzl}. In particular, for $c \geq 1$, the $1$-form symmetry is still given by $\Gamma^{(1)}=\BZ_2^{\fp/2}$ and the global form of the flavor symmetry is $\USp(2M\fp)/\BZ_2$; both are as in the case of $c=1$ discussed above.  The former is due to the fact that the monodromy of the Hitchin field around the irregular puncture for the theories with $c>1$ is the same as that for $c=1$ \cite{Bhardwaj:2021mzl}, whereas the latter can be seen from the mirror theory, which we will discuss in Section \ref{sec:3dmirrorzeromass}. The dictionary between line defects in the class $\CS$ theories and the aforementioned quantities was also provided in \cite{Bhardwaj:2021mzl}: 
\bi
\item the trapped electric (or magnetic) part $\CH_\CP^T$ of the defect group of the puncture $\CP$ is identified with $\Gamma^{(1)}$; 
\item the electric (or magnetic) part $\CH_\CP$ of the defect group of the puncture $\CP$ is identified with $\cE$; and 
\item the subgroup $Z_\CP$ of surface defects of a 6d $\mathcal{N}=(2,0)$ theory that can end at puncture $\CP$ is identified with $\cZ$.  
\ei
Moreover, in \cite[(9.16)-(9.17)]{Bhardwaj:2021mzl}, it was pointed out that the $D_{\fp c}(\USp(2M \fp))$ theory bears the structure described by exact sequence \eref{exactsequence} which, due to the dictionary above, implies a \emph{non-trivial extension} between the 1-form symmetry and the flavor symmetry and hence the \emph{2-group symmetry} for any $c$ such that $\GCD(c, 2M)=1$.\footnote{If we draw the Newton polygon for $D_{\fp}(\USp(2M \fp))$, we will find that it is bounded by a straight line such that the points $(0,2M\fp)$, $(1, 2M\fp-2M)$, $(2, 2M\fp-4M)$, $\ldots$,$(\fp-1, 2M)$, $(\fp, 0)$ lie on this straight line.  Since the first coordinate increases by $1$ and the second coordinate decreases by $2M$.  We take $q=2M$ in the notation of \cite{Bhardwaj:2021mzl}. We can choose $c>1$ which is co-prime to $q=2M$ and shift the aforementioned points, except for $(0,2M \fp)$, in the horizontal direction, so that we have  $(0,2M\fp)$, $(c, 2M\fp-2M)$, $(2c, 2M\fp-4M)$, $\ldots$, $((\fp-1)c, 2M)$, $(\fp c, 0)$.}  Indeed, we will provide a justification for the identification of the exact sequences \eref{exactseq2} and \eref{exactsequence} below. Observe that the 2-group structure in general $D_{\fp c}(\USp(2M \fp))$ theory is identical to that of $c=1$.

\subsubsection{Identification of the exact sequences} 
%\subsubsection{Justification for the identification of \eref{exactseq2} and \eref{exactsequence}} 
\label{sec:justification}

In order to justify the identification of \eref{exactseq2} and \eref{exactsequence}, it is convenient to deal with the Pontryagin dual frame associated with the line defects instead of defect groups or 1-form symmetries. Specifically, the short exact sequence dual to \eref{exactseq2} is (see \cite[(4.25)]{Bhardwaj:2021mzl})
\bes{ \label{pontexact}
0 \,\, \rightarrow \,\,\hat{Z}_\CP \,\, \rightarrow \,\, \CW_\CP \,\, \rightarrow\,\, \CW^T_\CP \,\, \rightarrow\,\,0~,
}
where $\CW^T_\CP$ denotes the line defects trapped in the irregular puncture $\CP$ charged under 1-form symmetry; and $\hat{Z}_\CP$ denotes the surface operators of the 6d theory inserted along a loop enclosing $\CP$. We can instead take this loop and collapse it on top of the regular puncture. In this case, according to \cite{Bhardwaj:2021ojs}, these surface operators give rise to flavor Wilson lines for the flavor symmetry associated with the regular puncture due to Dirichlet boundary conditions. Indeed, $\hat{Z}_\CP$ can be identified with the group of flavor Wilson lines, modulo screening by the adjoint representation of flavor symmetry.

For the $D_{\fp c} (\USp(2M \fp))$ theory in question, the regular puncture is the twisted full puncture of the $D$-type theory, which gives rise to the $\usp(2M \fp)$ flavor symmetry algebra. As will be shown explicitly in Section \ref{sec:3dmirrorzeromass} using the 3d mirror theory,\footnote{See also \cite{Bhardwaj:2021ojs} for the computation of the charge of the local operator under the center of the flavor symmetry.} there is no local operator charged under the $\BZ_2$ center of $\USp(2M\fp)$ and so the global form of the flavor symmetry is $\USp(2M \fp)/\BZ_2$.  In other words, the flavor Wilson lines associated with the regular puncture are not screened by genuine local operators.  According to \cite{Bhardwaj:2021wif}, the absence of screening implies the absence of the junction local operators, which means that the line operators $\mathcal{W}_\CP$ in \eref{pontexact} are associated with the non-trivial extension $\CE$ in \eref{exactsequence}. 

\subsubsection{Dynamical consequences of the 2-group} \label{sec:dynconsequences}

We now look at some simple dynamical consequences of the 2-group symmetry, following the discussion of \cite{Lee:2021crt}. We start by considering the $\Spin(4n+2)$ gauge theory with $N_f$ flavors of hypermultiplets in the vector representation, which we shall call it $\mathcal{T}$ for convenience. Although $\Spin(4n+2)$ has a $\BZ_4$ center, $\mathcal{T}$ has a $\Gamma^{(1)} = \mathbb{Z}_2$ 1-form symmetry due to the screening effect of the hypermultiplet in the vector representation \cite{Lee:2021crt}. Moreover, $\CT$ has a 2-group symmetry that is an $\CE = \BZ_4$ extension of the $\Gamma^{(1)} = \mathbb{Z}_2$ 1-form symmetry and the quotient $\CZ=\BZ_2$ of the flavor group $\USp(2N_f)/\BZ_2$. The fact that $\CE=\BZ_4$ can be seen as follows (see \cite[Section 2.4]{Apruzzi:2021mlh}). Let us consider the combined gauge-flavor center of $\Spin(4n+2)$ and $\USp(2N_f)$, namely $\BZ_4 \times \BZ_2$. The element $\omega=(e^{i 2\pi/4}, e^{i 2\pi/2}) = (i, -1)$ of $\BZ_4 \times \BZ_2$ acts trivially on the fundamental hypermultiplets, due to the fact that each element in $\omega$ acts on the latter as $(-1)$ and so the combined gauge-flavor action is $(-1)(-1)=+1$, which is trivial. Since $\omega^2 = (-1, 1)$ is not an identity but $\omega^4 = (1,1)$ is, we see that $\omega$ generates $\CE=\BZ_4$.

Let us now take four copies of $\CT$ and then gauge the diagonal $\USp(2N_f)$ symmetry. The resulting star-shaped quiver, which we will refer to as $\CS$, is\footnote{Note that for a sufficiently large $N_f$ the $\USp(2N_f)$ gauge group may be infrared free in 4d. This does not affect the subsequent discussion, since we are interested only in the global form of the gauge groups in the quiver.}
\bes{
\begin{tikzpicture}[baseline=0,font=\footnotesize]
\node[draw=none] (c) at (0,0) {$\USp(2N_f)$};
\node[draw=none] (n1) at (-1.5,1.5) {$\Spin(4n+2)$};
\node[draw=none] (n2) at (1.5,1.5) {$\Spin(4n+2)$};
\node[draw=none] (n3) at (1.5,-1.5) {$\Spin(4n+2)$};
\node[draw=none] (n4) at (-1.5,-1.5) {$\Spin(4n+2)$};
\draw (c)--(n1);
\draw (c)--(n2);
\draw (c)--(n3);
\draw (c)--(n4);
\end{tikzpicture}
}
This theory has a $\BZ_4 \times \BZ_2^3$ electric 1-form symmetry.\footnote{We thank the anonymous referee of the JHEP for pointing this out to us and for providing us with the argument.} This can be seen by a similar argument as above. We consider the element $\omega=(i,i,i,i,-1)$ of $\BZ_4 \times \BZ_4 \times \BZ_4 \times \BZ_4 \times \BZ_2$, which is the center of four $\Spin(4n+2)$ gauge groups and one $\USp(2N_f)$ gauge group. As discussed above, $\omega$ acts trivially on the $2N_f$ bifundamental hypermultiplets and it gives rise to a $\BZ_4$ factor in the 1-form symmetry. We can consider $\sigma_1 = (-1,1,1,1,1)$, $\sigma_2 = (1,-1,1,1,1)$, $\sigma_3 = (1,1,-1,1,1)$ and $\sigma_4 = (1,1,1,-1,1)$; each acts trivially on the bifundamental hypermultiplets and each generates a $\BZ_2$ factor of the 1-form symmetry. However, since $\prod_{i=1}^4 \sigma_i = \omega^2$, we see that the $\BZ_2$ diagonal subgroup of $\BZ_2^4$ is a subgroup of $\BZ_4$. We thus have a $\BZ_4 \times \BZ_2^3$ electric 1-form symmetry, as claimed.  Note that if we replaced each of the  $\Spin(4n+2)$ gauge group by $\Spin(4n)$, each copy of $\CT$ would have no 2-group symmetry and theory $\CS$ would have a $\BZ_2^5$ electric 1-form symmetry.  The presence of the $\BZ_4$ factor is therefore a crucial consequence of the 2-group symmetry of $\CT$.

Let us now gauge the whole 1-form symmetry, namely $\BZ_4 \times \BZ_2^3$, of $\CS$ and denote the resulting theory as $\CS'$.  The gauge group of $\CS'$ is
\begin{equation}
\begin{array}{*3{>{\displaystyle}l}p{5cm}}
& \frac{\Spin(4n+2) \times \Spin(4n+2) \times \Spin(4n+2) \times \Spin(4n+2) \times \USp(2N_f)}{\BZ_4 \times \BZ_2^3} & = \\
 = & \frac{\Spin(4n+2)/\BZ_2 \times \Spin(4n+2) \times \Spin(4n+2) \times \Spin(4n+2) \times\USp(2N_f)}{\BZ_2 \times \BZ_2^3} & = \\
=  & \frac{\SO(4n+2) \times \SO(4n+2) \times \SO(4n+2) \times \SO(4n+2) \times\USp(2N_f)}{\BZ_2 }~,
\end{array}
\end{equation}
where in the first equality we have used \cite[(4.19)]{Bhardwaj:2021ojs}.  Theory $\CS'$ can therefore be represented by the quiver 
\bes{ \label{starUSp8modZ2}
\begin{tikzpicture}[baseline=0,font=\footnotesize]
\node[draw=none] (c) at (0,0) {$\USp(2N_f)$};
\node[draw=none] (x) at (2.5,0) {$/\BZ_2$};
\node[draw=none] (n1) at (-1.5,1.5) {$\SO(4n+2)$};
\node[draw=none] (n2) at (1.5,1.5) {$\SO(4n+2)$};
\node[draw=none] (n3) at (1.5,-1.5) {$\SO(4n+2)$};
\node[draw=none] (n4) at (-1.5,-1.5) {$\SO(4n+2)$};
\draw (c)--(n1);
\draw (c)--(n2);
\draw (c)--(n3);
\draw (c)--(n4);
\end{tikzpicture}
}
where $/\BZ_2$ denotes the overall $\BZ_2$ quotient. If $\CS$ is regarded as a 4d $\CN=2$ gauge theory, then theory $\CS'$ has a $\BZ_4 \times \BZ_2^3$ magnetic 1-form symmetry.

It is also instructive to view $\CS$ as a 3d $\CN=4$ gauge theory.  In which case, gauging the $\BZ_4 \times \BZ_2^3$ 1-form symmetry of $\CS$ leads to a $\BZ_4 \times \BZ_2^3$ 0-form symmetry of $\CS'$. A natural question is how the latter arises from the perspective from the 3d supersymmetric index \cite{Bhattacharya:2008zy,Bhattacharya:2008bja, Kim:2009wb,Imamura:2011su, Kapustin:2011jm, Dimofte:2011py}.\footnote{Here we follow the convention of \cite{Aharony:2013kma}.} According to \cite[Section 6.1]{Aharony:2013kma}, each of the four $\SO(4n+2)$ gauge groups gives rise to a factor $\zeta_i^{\sum_{j=1}^{2n+1} m^{(i)}_j}$, where $i=1, 2, 3, 4$; $\zeta_i$ is the fugacity of the $\BZ_2$ 0-form magnetic symmetry of the $i$-th $\SO(4n+2)$ gauge group such that $\zeta_i^2=1$; and $m^{(i)}_j$ are the magnetic fluxes of the $i$-th $\SO(4n+2)$ gauge group. Let $n_l$ (with $l=1, \ldots, N_f$) be magnetic fluxes for $\USp(2N_f)$.  Modding out by an overall $\BZ_2$ means that all $m^{(i)}_j$ and $n_l$ are integers, or all $m^{(i)}_j$ and $n_l$ are half-odd-integers. We have to sum over all of these magnetic fluxes in the index.  Since each $\zeta_i$ is $\BZ_2$-valued, in the half-odd-integer case, we have $w \equiv \prod_{i=1}^4 \zeta_i^{\pm 1/2}$ in the index; this is indeed a manifestation of the $\BZ_4$ factor in the $0$-th form symmetry.\footnote{This should be contrasted with the case of the $\SO(4n)$ gauge group which contributes $\zeta^{\sum_{j=1}^{2n} m_j}$. In this case, half-odd-integral magnetic fluxes $m_j$ lead to an integer power of $\zeta$, not a half-odd-integer power as for the case of $\SO(4n+2)$.} Note that $w^2 = \prod_{i=1}^4 \zeta_i$, and so the diagonal $\BZ_2$ subgroup of $\BZ_2^4$ that corresponds to the fugacity $\prod_{i=1}^4 \zeta_i$ is actually a subgroup of the $\BZ_4$ symmetry.

Let us demonstrate this explicitly in the special case of $n=0$ and $N_f=1$. Here $\CT$ is the $\Spin(2)$ gauge theory with 1 flavor of hypermultiplets in the vector representation, which is equivalent to the $\U(1)$ gauge theory with 2 hypermultiplets of charge 2 (see \cite{Mekareeya:2022spm}). Gauging the $\BZ_2$ 1-form symmetry, we see that the $\SO(2)$ gauge theory with 1 flavor of hypermultiplets in the vector representation is identified with the $\U(1)$ gauge theory with 2 hypermultiplets of charge 1, which flows to the $T[\SU(2)]$ SCFT \cite{Gaiotto:2008ak}. The $\BZ_2$ 0-form magnetic symmetry of the $\SO(2)$ gauge theory should be treated as a subgroup of the $\SO(3)$ enhanced topological symmetry of $T[\SU(2)]$. Theory $\CS'$ is thus the well-known star-shaped mirror theory of the 3d $\CN=4$ $\SU(2)$ gauge theory with $4$ hypermultiplets in the fundamental representation \cite{Benini:2010uu}.  The index of $\CS'$ is given by (see \eg~ \cite{Razamat:2014pta, Razamat:2019sea})
\bes{ \label{indSp}
\CI_{\CS'}(\zeta_{1,2,3,4} ; x)  = \frac{1}{2} \sum_{\epsilon=0}^1 \sum_{n \in \BZ+\frac{1}{2}\epsilon} \oint \frac{d z}{2\pi i z}  \CZ^{\USp(2)}_{\text{vec}} (z,n) \prod_{i=1}^4 \CI^\epsilon_{T[\SU(2)]} (\zeta_i,p_i =0|z,n;x)~, 
}
where the $\USp(2)$ vector multiplet contribution is
\bes{
\CZ^{\USp(2)}_{\text{vec}} (z,n) = x^{-2|n|} \prod_{s=\pm 1}(1- (-1)^{2n} x^{2|n|} z^{2s})~,
}
the index of the $T[\SU(2)]$ theory, computed from the $\U(1)$ gauge theory with 2 hypermultiplets of charge 1, is
\bes{
\CI^\epsilon_{T[\SU(2)]} (\zeta,p|z,n;x) &= \sum_{m \in \BZ+\frac{1}{2}\epsilon} \zeta^m \oint \frac{d z}{2\pi i z} z^p\, \prod_{s=\pm 1} \CI^{1/2}_\chi ((z f)^s; s(m+n);x) \times \\
& \qquad\qquad\qquad \CI^{1/2}_\chi ((z^{-1} f)^s; s(-m+n);x)
}
such that for $\epsilon=0$, $n$ is integral and for $\epsilon=1$, $n$ is half-odd-integral, and the contribution of the chiral multiplet of $R$-charge $R$ is
\bes{
I^R_{\chi}(z, m;x) = \left( x^{1-R} z^{-1} \right)^{|m|/2} \prod_{j=0}^\infty \frac{1-(-1)^m z^{-1} x^{|m|+2-R+2j}}{1-(-1)^m z\,  x^{|m|+R+2j}}~.
}
For convenience, let us discuss the result of \eref{indSp} up to order $x$:
\bes{
\CI_{\CS'}(\zeta_{1,2,3,4} ; x)  = 1+ x \left( 4+ \sum_{i=1}^4 (\zeta_i + \zeta_i^{-1}) + \sum_{\substack{s_1, \cdots, s_4 \\  = \pm 1} }\prod^4_{i=1} \zeta^{\frac{1}{2} s_i}_i \right) + \ldots~.
}
The terms at order $x$ indeed correspond to the character of the adjoint representation of $\so(8)$, which is the flavor symmetry algebra of the $\SU(2)$ gauge theory with 4 flavors. The 16 terms involving the half-odd-integer powers of $\zeta_i$ arise from $\epsilon=1$ which corresponds to the half-odd-integral magnetic fluxes, whereas the remaining 12 terms arise from the integral magnetic fluxes.

This argument can be generalized easily to star-shaped quiver theories containing $\CT = D_{\fp c}(\USp(2M\fp))$, with $M$ even, as a building block.

\subsubsection{\texorpdfstring{$M$}{} odd}

For $M$ odd and $c=1$, all the $D$-type gauge groups in \eref{LagDpUSp2Mp} have even rank, and such a theory does not have a non-trivial 2-group structure \cite{Lee:2021crt, Apruzzi:2021mlh}. Nevertheless, it has a non-trivial $\BZ^{\fp/2}_2$ 1-form symmetry. Using the argument from Section \ref{sec:bootstrap}, we conclude that all $D_{\fp c}(\USp(2M \fp))$ theories, with $\fp$ even, $M$ odd and $\GCD(c, 2M)=1$, have a $\BZ^{\fp/2}_2$ 1-form symmetry but do not have a 2-group structure.

\subsubsection{3d mirror theories} \label{sec:3dmirrorzeromass}

Let us now discuss the 3d mirror theories for the $D_{\fp c}(\USp(2M\fp))$ and $C^{2N}[\fp c]$ theories.

Assuming that $c \geq 2M$ and that $c$ is coprime to $2M$, we can close the full puncture in the $D_{\fp c}(\USp(2M\fp))$ theory and obtain the $C^{2N}[\fp c]$ theory. The latter is {\it non-Higgsable} with rank\footnote{Half of the Milnor number for the $C^{2N}[\kappa]$ theory is $\frac{1}{2}\left[(\kappa-2N+1)N +\kappa-2N \right]$. If this is a non-Higgsable theory, the rank for such a theory is exactly equal to this quantity.} 
\bes{
r_0 \equiv \frac{1}{2}\fp [c-M+(c-2M)M \fp]~,
}
and with a 1-form symmetry $\BZ_2^{\fp/2}$.  This leads to the following conclusions: 
\bi
\item The mirror of the $C^{2Mp}[\fp c]$ theory, with $\fp$ even and $\GCD(c, 2M)=1$, is a theory of $r_0$ free hypermultiplets, which may be subject to a $\BZ^{\fp/2}_2$ discrete gauging.
\item The mirror theory for $D_{\fp c}(\USp(2M\fp))$, with $\fp$ even, $\GCD(c, 2M)=1$ and $c \geq 2M$, is $r_0$ free hypermultiplets (which may be subject to a $\BZ^{\fp/2}_2$ discrete gauging) plus the $T[\SO(2M \fp+1)]$ theory,\footnote{We emphasize that the interacting part of the mirror theory does not depend on $c$; the information about $c$ is actually contained in the free sector.} whose quiver description of the theory is
\bes{ \label{quivTSO2Mpp1}
B_0 -C_1-B_1-C_2-B_2 - \cdots - &C_{Mp} - [B_{Mp}]
}
\ei
Since the generators of the Coulomb branch of the mirror theory $T[\SO(2M \fp+1)]$ of $D_{\fp c}(\USp(2M\fp))$ transform in the adjoint representation of $\USp(2M \fp)$, it follows that the Coulomb branch symmetry of the $T[\SO(2M \fp+1)]$ theory and hence the flavor symmetry of $D_{\fp c}(\USp(2M\fp))$ is $\USp(2M \fp)/\BZ_2$, due to the fact that there is no operator charged under the $\BZ_2$ center of $\USp(2M \fp)$.

\subsubsection{Partially closing the full puncture}

We can obtain several new interesting theories by studying the closure of the regular D-twisted puncture. Since the 1-form symmetry is invariant under the Higgs branch flow \cite{Carta:2022spy}, it follows that the 1-form symmetry before and after (partially) closing the punctures are the same. We demonstrate this explicitly in each theory discussed below using the description that involves weakly gauging. We also discuss the presence or absence of the 2-group structure in each theory. 

\begin{example}{\textbf{$D_6(\USp(8))$ with $[1^8]$ being partially closed to $[6,1^2]$ or $[6,2]$}}\\

Let us consider for instance the $D_6(\USp(8))$ theory.  Note that this can be bootstrapped from the $D_2(\USp(8))$ theory, which admits the following Lagrangian description:
\bes{
D_3-[C_4]~.
}
Suppose that $D_3$ is chosen to be $\Spin(6)$, then this theory has an electric $\BZ_2$ 1-form symmetry and a $2$-group structure between the 1-form symmetry and the $\USp(8)/\BZ_2$ flavor symmetry.\footnote{If, instead, $D_3$ is chosen to be $\SO(6)$, or equivalently the electric $\BZ_2$ 1-form symmetry of the aforementioned $\Spin(6)$ theory is gauged, then this theory has a magnetic $\BZ_2$ 1-form symmetry and no $2$-group structure.  Instead, there is a mixed anomaly between the $\USp(8)/\BZ_2$ flavor symmetry \cite{Lee:2021crt} and this magnetic 1-form symmetry.}  By the argument in the previous section, the $D_6(\USp(8))$ also has the $\BZ_2$ 1-form symmetry.  The Coulomb branch spectrum of $D_6(\USp(8))$ consists of operators of dimension 
\be\label{speccb} \text{CB} = \left\{\frac{4}{3},\frac{4}{3},\frac{5}{3},2,\frac{8}{3},\frac{8}{3},3,\frac{10}{3},4,\frac{13}{3},\frac{14}{3},\frac{16}{3},\frac{20}{3}\right\}\fstop\ee 
The easiest way to derive this is to draw the Newton polygon of the $D_5$ twisted theory. This is obtained by plotting on the plane all the terms appearing in the spectral equation of the theory, which reads 
\be\label{speccc} {\color{blue} x^5}+{\color{blue} xz^p}+x^4P_4(z)+x^3P_3(z)+x^2P_2(z)+xP_1(z)+{\color{green}z\widetilde{P}^2(z)}=0.\ee 
In \eqref{speccc}, we marked in blue the leading terms of the singularity and in green the term associated with the Casimir of $D_5$ (the Pfaffian) that is not invariant under the outer-automorphism twist. Plotting these terms on the plane, we find:
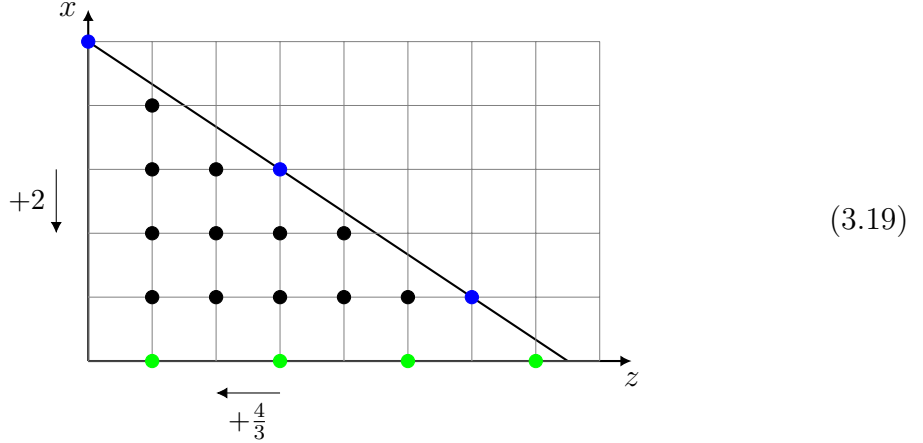
\bes{
\begin{tikzpicture}[baseline,scale=0.85]
\def\szp{5pt};
\draw[ligne,black] (0,5)--(7.5,0);
\draw[thick,->] (0,0)-- node[left, pos=1] {$x$} (0,5.5);
\draw[thick,->] (0,0)-- node[below, pos=1] {$z$} (8.5,0);
\draw[step=1cm,gray,very thin] (0,0) grid (8,5); 
\node[circle,draw=black,fill=black,inner sep=0pt,minimum size=\szp] at (1,1) {};
\node[circle,draw=black,fill=black,inner sep=0pt,minimum size=\szp] at (2,1) {};
\node[circle,draw=black,fill=black,inner sep=0pt,minimum size=\szp] at (3,1) {};
\node[circle,draw=black,fill=black,inner sep=0pt,minimum size=\szp] at (4,1) {};
\node[circle,draw=black,fill=black,inner sep=0pt,minimum size=\szp] at (5,1) {};
\node[circle,draw=black,fill=black,inner sep=0pt,minimum size=\szp] at (1,2) {};
\node[circle,draw=black,fill=black,inner sep=0pt,minimum size=\szp] at (2,2) {};
\node[circle,draw=black,fill=black,inner sep=0pt,minimum size=\szp] at (3,2) {};
\node[circle,draw=blue,fill=blue  ,inner sep=0pt,minimum size=\szp] at (3,3) {};
\node[circle,draw=blue,fill=blue  ,inner sep=0pt,minimum size=\szp] at (0,5) {};
\node[circle,draw=blue,fill=blue  ,inner sep=0pt,minimum size=\szp] at (6,1) {};
\node[circle,draw=black,fill=black,inner sep=0pt,minimum size=\szp] at (4,2) {};
\node[circle,draw=black,fill=black,inner sep=0pt,minimum size=\szp] at (1,3) {};
\node[circle,draw=black,fill=black,inner sep=0pt,minimum size=\szp] at (2,3) {};
\node[circle,draw=black,fill=black,inner sep=0pt,minimum size=\szp] at (1,4) {};
\node[circle,draw=green,fill=green,inner sep=0pt,minimum size=\szp] at (1,0) {};
\node[circle,draw=green,fill=green,inner sep=0pt,minimum size=\szp] at (3,0) {};
\node[circle,draw=green,fill=green,inner sep=0pt,minimum size=\szp] at (5,0) {};
\node[circle,draw=green,fill=green,inner sep=0pt,minimum size=\szp] at (7,0) {};
\draw[->] (-0.5,3) -- node[midway,left] {\small $+2$} (-0.5,2);
\draw[->] (3,-0.5) -- node[midway,below] {\small $+\frac{4}{3}$} (2,-0.5);
\end{tikzpicture}
}
The dots in the above picture denote CB operators and relevant couplings of the SCFT. 
The nodes in green indicate that the corresponding operator is actually the square of a chiral ring generator, and the nodes in blue denote the leading singular terms and the marginal coupling (since it lies on the diagonal). 
The scaling dimension of the various parameters can be determined as follows: The marginal coupling has indeed dimension zero; as we move downwards by one step the dimension increases by $2$ and moving to the left by one step increases the dimension by $\frac{4}{3}$. We should halve the dimension when we consider green nodes. One can easily check that following this procedure, we reproduce the CB spectrum \eqref{speccb}. 
Specifically, the term $P_4(z)$ in \eqref{speccc} is linear (since terms of the form $x^4z^k$ with $k>1$ are subleading with respect to $x^5$ and $xz^6$) and corresponds to the quadratic casimir of $D_5$. Besides the mass associated with the $\USp(8)$ global symmetry, it also describes a relevant coupling of dimension $\frac{2}{3}$ which corresponds to the black dot with coordinates $(1,4)$ in the plane. Analogously, $P_3(z)$ stands for the quartic casimir of $D_5$ and its coefficients correspond to the marginal coupling (the blue dot with coordinates $(3,3)$) and CB operators of dimension $\frac{4}{3}$ and $\frac{8}{3}$ respectively (the black dots with coordinates $(2,3)$ and $(1,3)$). $P_2(z)$ and $P_1(z)$ correspond to the casimir invariants of $D_5$ with degree 6 and 8 respectively. The pfaffian (the term in green in \eqref{speccc}) corresponds instead to $\sqrt{z}\widetilde{P}(z)$ and changes sign upon looping around the $z$-plane, as expected due to the outer-automorphism twist.

If we now close the $[1^8]$ puncture to $[6,1^2]$ we find, following the procedure described in \cite{Carta:2021whq} (see \cite{Chacaltana:2013oka} for details about $D$-twisted punctures), that the  IR theory has CB operators of dimension 
\be \text{CB}^{\text{IR}} = \left\{\frac{4}{3},\frac{4}{3},2,\frac{8}{3}\right\}\fstop\ee 
This is seen as follows: The pole orders of the full puncture $[1^8]$ are given by the sequence 
\be \left\{1,3,5,7,\frac{9}{2}\right\}\fstop\ee 
Those of the puncture $[6,1^2]$ are instead 
\be\label{poles} \left\{1,3,3,3,\frac{3}{2}\right\}\fstop\ee 
According to the procedure proposed in \cite{Carta:2021whq} we should then read off the CB spectrum again from the Newton polygon we have described before 
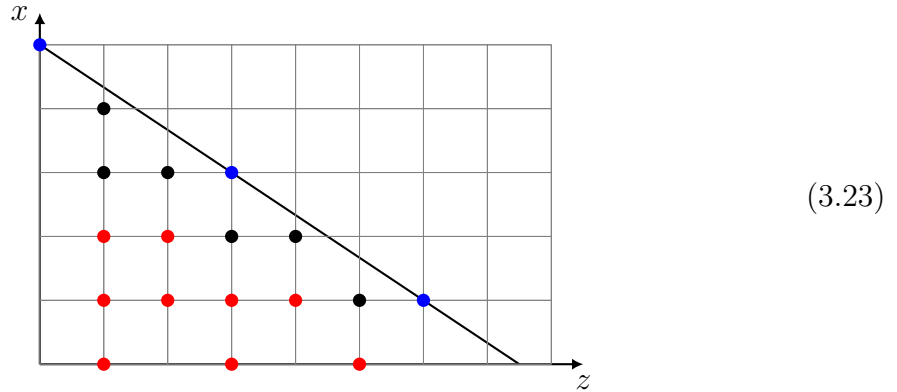
\bes{
\begin{tikzpicture}[baseline,scale=0.85]
\def\szp{5pt};
\draw[ligne,black] (0,5)--(7.5,0);
\draw[thick,->] (0,0)-- node[left, pos=1] {$x$} (0,5.5);
\draw[thick,->] (0,0)-- node[below, pos=1] {$z$} (8.5,0);
\draw[step=1cm,gray,very thin] (0,0) grid (8,5); 
\draw[step=1cm,gray,very thin] (0,0) grid (8,5); 
\node[circle,fill=red,inner sep=0pt,  minimum size=\szp] at (1,1) {};
\node[circle,fill=red,inner sep=0pt,  minimum size=\szp] at (2,1) {};
\node[circle,fill=red,inner sep=0pt,  minimum size=\szp] at (3,1) {};
\node[circle,fill=red,inner sep=0pt,  minimum size=\szp] at (4,1) {};
\node[circle,fill=black,inner sep=0pt,minimum size=\szp] at (5,1) {};
\node[circle,fill=red,inner sep=0pt,  minimum size=\szp] at (1,2) {};
\node[circle,fill=red,inner sep=0pt,  minimum size=\szp] at (2,2) {};
\node[circle,fill=black,inner sep=0pt,minimum size=\szp] at (3,2) {};
\node[circle,fill=blue,inner sep=0pt, minimum size=\szp] at (3,3) {};
\node[circle,fill=blue,inner sep=0pt, minimum size=\szp] at (0,5) {};
\node[circle,fill=blue,inner sep=0pt, minimum size=\szp] at (6,1) {};
\node[circle,fill=black,inner sep=0pt,minimum size=\szp] at (4,2) {};
\node[circle,fill=black,inner sep=0pt,minimum size=\szp] at (1,3) {};
\node[circle,fill=black,inner sep=0pt,minimum size=\szp] at (2,3) {};
\node[circle,fill=black,inner sep=0pt,minimum size=\szp] at (1,4) {};
\node[circle,fill=red,inner sep=0pt,  minimum size=\szp] at (1,0) {};
\node[circle,fill=red,inner sep=0pt,  minimum size=\szp] at (3,0) {};
\node[circle,fill=red,inner sep=0pt,  minimum size=\szp] at (5,0) {};
\end{tikzpicture}
}
but with the red nodes omitted. Since the puncture $[6,1^2]$ does not produce any constraints, this is the final result. The conclusion is different for the puncture $[6,2]$ since the pole structure is the same as in \eqref{poles} but there is a constraint of A-type, implying that the parameter of dimension $\frac{8}{3}$ is actually the square of a CB operator, whose dimension is therefore $\frac{4}{3}$.

We can actually describe this theory in detail, since it coincides with a $\su(2)$ vector multiplet coupled both to the rank-1 $H_1$, which is also known as the $(A_1,A_3)$ theory, and the rank-2 $H_1$ theories in the F-theory terminology. The Higgs branch of the rank-2 $H_1$ theory is isomorphic to the reduced moduli space of two $\SU(2)$ instantons on $\BC^2$, whose isometry is $\su(2)_{\BC^2} \times \su(2)_{\text{inst}}$, where $\su(2)_{\BC^2}$ is inherited from the isometry of $\BC^2$ and $\su(2)_{\text{inst}}$ is the instanton gauge group itself. On the other hand, the Higgs branch of the $(A_1, A_3)$ theory is isomorphic to the reduced moduli space of one $\SU(2)$ instanton, namely $\BC^2/\BZ_2$, whose isometry is $\su(2)_{\text{inst}}$.  Gauging the common $\su(2)_{\text{inst}}$ gauge algebra, we see that the theory in question can be described pictorially as follows:\footnote{One can easily check that the beta function for the $\su(2)_{\text{inst}}$ gauge algebra vanishes exploiting the results of \cite{Aharony:2007dj} about the flavor central charge of instanton theories.}
\bes{ \label{closeto611}
\begin{tikzpicture}[baseline]
\node[rectangle] (rank2) at (-2,0) {$\text{rank-2 $H_1$}$};
\node[rectangle] (rank1) at (2,0) {$(A_1, A_3)$};
\node[draw=none] (flv) at (-6,0) {};
\draw[red,solid,thick] (rank2) to node[above]{$\scriptsize{\su(2)_{\text{inst}}}$} (rank1);
\draw[blue,solid,thick] (rank2) to node[above]{$\scriptsize{\su(2)_{\BC^2}}$} (flv);
\end{tikzpicture}
}
where the red line denotes the gauging of the common $\su(2)_{\text{inst}}$ symmetry algebra of the two theories, and the semi-infinite blue line on the left denotes the $\su(2)_{\BC^2}$ flavor symmetry algebra. Using the information from \cite[(3.11)]{Benvenuti:2010pq} and \cite[(3.14)-(3.15)]{Hanany:2012dm}, we compute the Higgs branch Hilbert series of theory \eref{closeto611} and find that the Higgs branch is precisely isomorphic to $\BC^2/\BZ_2$.\footnote{The quaternionic Higgs branch dimension of \eref{closeto611} is indeed $(2\times 2-1)+1-3=1$. Explicitly, the Higgs branch Hilbert series is
\bes{ \nn
 \oint_{|z|=1} \frac{1-z^2}{2\pi i z} \PE\left[ -t^2 \chi^{\su(2)}_{[2]} \right] \times \tilde{g}_{2, \SU(2)}(t, x, z)  \times \left(\sum_{k=0}^\infty \chi^{\su(2)}_{[2k]} t^{2k} \right) = \PE \left[ \chi^{\su(2)}_{[2]}t^2-t^4\right]~,
}
where the function $\tilde{g}_{2, \SU(2)}(t, x, z)$ is given by \cite[(3.14)]{Hanany:2012dm}.
}  The global form of the flavor symmetry, corresponding to the blue line, is therefore $\SO(3)_{\BC^2}$.  Let us now determine the 1-form symmetry of theory \eref{closeto611}.  As described in \cite[(3.15)]{Hanany:2012dm}, there are two generators of the two $\SU(2)$ instanton moduli space: one (the moment map) in the $[2;0]+[0;2]$ representation and the other in the $[1;2]$ representation of $\su(2)_{\BC^2} \times \su(2)_{\text{inst}}$. The global form of the flavor symmetry of the rank-2 $H_1$ theory is $\SU(2)_{\BC^2} \times \SO(3)_{\text{inst}}$.  On the other hand, the Higgs branch of the $(A_1, A_3)$ theory, which is $\BC^2/\BZ_2$, is generated by a single generator in the adjoint representation of the $\su(2)_{\text{inst}}$ and so the global form of the flavor symmetry of $(A_1, A_3)$ is $\SO(3)_{\text{inst}}$. Suppose that the gauge group corresponding to $\su(2)_{\text{inst}}$ denoted by the red line in \eref{closeto611} is chosen to be $\SU(2)_{\text{inst}}$, then the theory in \eref{closeto611} has an electric 1-form $\BZ_2$ symmetry, arising from the center of $\SU(2)_{\text{inst}}$.\footnote{On the other hand, if $\su(2)_{\text{inst}}$ is chosen to be $\SO(3)_{\text{inst}}$, or equivalently the electric $1$-form symmetry of the aforementioned $\SU(2)_{\text{inst}}$ theory is gauged, then the theory has a magnetic 1-form $\BZ_2$ symmetry.} Note that the presence of the $\BZ_2$ 1-form symmetry is the same as that of the original $D_6(\USp(8))$ theory, which is in agreement with the fact that the 1-form symmetry is invariant under the Higgs branch flow \cite{Carta:2022spy}. It is worth emphasizing that gauging the $\SU(2)_{\text{inst}}$ symmetry of the rank-2 $H_1$ theory and coupling it to the $(A_1, A_3)$ theory changes the $\SU(2)_{\BC^2}$ global symmetry of the rank-2 $H_1$ theory to the $\SO(3)_{\BC^2}$ symmetry in \eref{closeto611}. In other words, the $\BZ_2$ electric 1-form symmetry, which arises from the center of the $\SU(2)_{\text{inst}}$ gauge group, comes hand-in-hand with the 2nd Stiefel-Whitney class $w_2$ that is an obstruction for lifting the $\SO(3)_{\BC^2}$ bundles to the $\SU(2)_{\BC^2}$ bundles in \eref{closeto611}. It is therefore expected that there is a 2-group symmetry between such a $\BZ_2$ electric 1-form symmetry and the $\SO(3)_{\BC^2}$ flavor symmetry.\footnote{On the other hand, if this $\BZ_2$ electric 1-form symmetry is gauged, or equivalently $\su(2)_{\text{inst}}$ in \eref{closeto611} is chosen to be $\SO(3)_{\text{inst}}$, then it is expected that there is a mixed anomaly between the new $\BZ_2$ magnetic 1-form symmetry and the $\SO(3)_{\BC^2}$ flavor symmetry \cite{Tachikawa:2017gyf, Lee:2021crt}.}

By giving a nilpotent VEV to the corresponding moment map of $\SU(2)_{\BC^2}$, we further close the puncture to $[6,2]$, leading to the spectrum \be \text{CB}^{\text{IR}} = \left\{\frac{4}{3},\frac{4}{3},\frac{4}{3},2\right\}\fstop\ee 
This corresponds to the $\SU(2)$ vector multiplet coupled to three copies of the rank-1 $H_1$ theory. The result turns out to be the $(A_2,D_4)$ theory\footnote{As pointed out in  \cite[Section 5.3]{Carta:2021whq}, reduction to 3d of this 4d theory gives a star-shaped quiver arising from gauging the common $\su(2)$ symmetry of three copies of the 3d $\CN=4$ $\U(1)$ gauge theory with two hypermultiplets of charge 1. If the gauge group corresponding to this $\su(2)$ is chosen to be $\SU(2)/\BZ_2$, then the mirror theory is the theory of free four hypermultiplets (also known as the $T_2$ theory), whereas if it is chosen to be $\SU(2)$, then the mirror theory is the $\BZ_2$ discrete quotient of the $T_2$ theory. The two possible choices of such a gauge group reflects the presence of the $\BZ_2$ 1-form symmetry of the 4d theory in question \cite{Carta:2021whq, Carta:2022spy}.} (see \cite[Figure 1]{Buican:2016arp} and \cite[(3.2)]{Closset:2020afy}). This is indeed expected, since it is well known that activating a nilpotent VEV for the $\SU(2)$ moment map higgses the two-instanton theory to two copies of the one-instanton theory \cite{Beem:2019snk}. We can easily see that the IR theories we have just described reproduce the 1-form $\mathbb{Z}_2$ symmetry of the UV theory, since the matter sectors have trivial charge under the center of the $\SU(2)$ gauge group. 

\end{example}

%\newpage % zzzz

\begin{example}{\textbf{$D_4(\USp(8))$ with $[1^8]$ being partially closed to $[4,1^4]$ or $[4,2,1^2]$}}\\

Another interesting example is given by the $D_4(\USp(8))$ theory, which is Lagrangian and coincides with the linear quiver 
\be \label{D4USp8} D_2-C_2-D_4-[C_4]\fstop\ee
Let us take each $D_{2n}$ node to be a $\Spin(2n)$ gauge group. This theory has a $\BZ_2^2$ 1-form symmetry.  The global form of the flavor symmetry is $\USp(8)/\BZ_2$, where the generators of the Higgs branch transform under the adjoint representation of $\USp(8)$ and so there is no operator charged under the $\BZ_2$ center of $\USp(8)$. We expect that there is no $2$-group symmetry, since each gauge group of the $D$-type has even rank \cite{Lee:2021crt, Apruzzi:2021mlh}.

If we close the full puncture $[1^{8}]$ to $[4,1^4]$ the theory is higgsed to 
\bes{ \label{closeto414} 
\begin{tikzpicture}[baseline]
	\node (g1) [] at (2,0) {$\USp(4)$};
	\node (g2) [] at (0,0) {$\Spin(4)$};
	\node (fAS) at (3.7,0) {$[2]$};
	\draw [snake it]  (g1)--(fAS);
	\draw (3.1,0.25) node {$\scriptscriptstyle{\Lambda^2}$};
	\draw (g1)--(g2);
\end{tikzpicture}
}
where the wiggly line on the right denotes two hypermultiplets in the antisymmetric representation of the $\USp(4)$ gauge group. The global form of the flavor symmetry of this theory is $\USp(4)/\BZ_2$, since the generators of the Higgs branch transform in the adjoint representation of $\mathfrak{usp}(4)$ flavor algebra and so there are no operators charged under the $\BZ_2$ center. Since the 1-form symmetry is invariant under the Higgs branch flow \cite{Carta:2022spy}, we expect that this theory also has the same 1-form symmetry as that of \eref{D4USp8}, namely $\BZ_2^2$. This can be seen from the Lagrangian description as follows: the $\Spin(4)$ group has a $\BZ_2^2$ center, the $\USp(4)$ group has a $\BZ_2$ center, and the bifundamental hypermultiplet screens the Wilson line in the $(\mathbf{4}; \mathbf{4})$ representation of $\Spin(4) \times \USp(4)$ and so the Wilson lines that are not screened are those in the representations $(\mathbf{2_s}; \mathbf{4})$ and $(\mathbf{2_c}; \mathbf{4})$; and they are those charged under the aforementioned $\BZ_2^2$ 1-form symmetry. The presence of hypermultiplets in the antisymmetric representation of $\USp(4)$ does not affect such a 1-form symmetry. Since $\Spin(4)$ is a $D$-type gauge group of an even rank, we do not expect the presence of the $2$-group symmetry in this theory.

If we turn on a VEV for one of the antisymmetric hypermultiplets in \eref{closeto414} (corresponding to the minimal nilpotent VEV for the moment map of the $\USp(4)$ global symmetry), we further close the puncture to $[4,2,1^2]$ and the corresponding field theory coincides with the Class $\mathcal{S}$ theory of type $A_1$ on a genus-$2$ surface with one regular puncture: 
\bes{\label{genus2}
\begin{tikzpicture}[baseline]
\node[mark size=15pt, rotate=-90] at (6,0) {\pgfuseplotmark{triangle}};
\node[minimum height=.8cm, minimum width=1cm] (L1) at (6,0) {\scriptsize $T_2$}; 
\path[every node/.style={font=\sffamily\small,
  		fill=white,inner sep=1pt}]
(L1) edge [loop, out=120, in=240, looseness=8] (L1);
\node[draw, circle, fill=white] (node6) at (4.8,0) {\begin{tiny} $\SU(2)$\end{tiny}};
\node[draw, circle, fill=white] (node8) at (7.5,0) {\begin{tiny}$\SU(2)$\end{tiny}};
\node[mark size=15pt] at (9,0.25) {\pgfuseplotmark{triangle}};
\node[minimum height=.8cm, minimum width=1cm] (L3) at (9,0.25) {\scriptsize $T_2$};
\node[draw, circle, fill=white] (node9) at (10.5,0) {\begin{tiny}$\SU(2)$\end{tiny}};
\node[mark size=15pt, rotate=90] at (12,0) {\pgfuseplotmark{triangle}};
\node[minimum height=.8cm,  minimum width=1cm] (L2) at (12,0) {\scriptsize $T_2$}; 
\path[every node/.style={font=\sffamily\small,
  		fill=white,inner sep=1pt}]
(L2) edge [loop, out=60, in=300, looseness=8] (L2);
\node[draw, circle, fill=white] (node6) at (13.2,0) {\begin{tiny} $\SU(2)$\end{tiny}};
\draw[thick]  (L1) to (node8) to (node9) to (L2);
\end{tikzpicture}
}
where the triangles in Figure \eqref{genus2} denote the $T_2$ theories of \cite{Gaiotto:2009we}, namely a half-hypermultiplet in the trifundamental of $\SU(2)^3$. Since there are two $\SU(2)$ gauge groups in the Lagrangian description, we see that this theory has a $\BZ_2^2$ 1-form symmetry, which is identical to that of the aforementioned theories. This is indeed in agreement with the observation that the 1-form symmetry is invariant under the Higgs branch flow \cite{Carta:2022spy}. The Higgs branch Hilbert series of this theory is given by \cite[(7.8)]{Hanany:2010qu}, with $\chi=3$, where it was pointed out that the Higgs branch has two generators: one (the moment map) in the adjoint representation and the other in the fundamental representation of the $\su(2)$ flavor symmetry algebra. Due to the presence of the latter, the global form of the flavor symmetry of the theory is $\SU(2)$, and not $\SU(2)/\BZ_2$.  As a result, this theory does not have a $2$-group symmetry.

\end{example}

\subsection{One mass parameter}

Every $D_p(\USp(2N))$ theory with one mass parameter, excluding the Casimirs of $\USp(2N)$, must satisfy the following conditions: If $N=1$, then $p$ can be any integer greater than 1; if $N$ is even, then $p$ must be odd; and if $N$ is odd and $N\geq3$, then $p$ must not be divisible by any $x$ such that $x>2$, $x$ is even, and $x$ divides $2N$. Such theories can be divided into two main subclasses as follows:
\begin{enumerate}
    \item One subclass of the theories with one mass parameter contains the $D_{\fp}(\USp(2M\fp))$ theory with $\fp$ odd, which admits the following Lagrangian description\footnote{Note that, for $\fp=1$, this theory, namely $D_{1}(\USp(2M))$, is just a theory of $M$ free hypermultiplets.} \cite[(A.9)]{Carta:2021dyx}
\bes{ \label{DpUSp2Mp}
[C_{M \fp}]-D_{M \fp-(M-1)}-\cdots-C_{3M}-D_{2M+1}-C_M-[D_1]
}
Using the similar argument as in \cite[(3.9)-(3.12)]{DelZotto:2020esg}, we see that the 1-form symmetry of this theory is $\BZ_2^{(\fp-1)/2}$, where $(\fp-1)/2$ is the number of the $D$-type gauge groups (all of their ranks are odd). Moreover, from the quiver, it is clear there are operators transforming in the representation $[\mathbf{2M \fp}; \pm 1]$ of the flavor symmetry algebra $\usp(2M \fp) \times \mathfrak{u}(1)$. The global form of the flavor symmetry is therefore $\USp(2 M \fp) \times \U(1)$, and not $\frac{\USp(2 M \fp) \times \U(1)}{\BZ_2}$.  Since the quotient of the flavor symmetry is trivial, there is no obstruction class analogous to the 2nd Stiefel-Whitney class, and so \eref{DpUSp2Mp} has no 2-group symmetry. 

Let us now bootstrap theory \eref{DpUSp2Mp}. We obtain a family of the $D_{\fp c}(\USp(2M\fp))$ theories with $\fp$ odd and $c$ coprime to $q\equiv 2M$.  Note that this is in fact equivalent to the family of the $D_p(\USp(2N))$ theories, with $p$ odd.  Since this process preserves the 1-form symmetry and the global form of the flavor symmetry, we deduce that all of such theories have a $\BZ_2^{(\fp-1)/2}$ 1-form symmetry and trivial 2-group symmetry. 

\item The other subclass of the $D_p(\USp(2N))$ theories with one mass parameter does not contain a theory with a 4d $\CN=2$ Lagrangian description; in other words, it cannot be bootstrapped from \eref{DpUSp2Mp}.\footnote{This subclass is characterized by $D_{\fp}(\USp(2))$ with $\fp$ even and $D_{c \,\GCD(2N,\fp)}(\USp(2N))$ with odd $N\geq 3$ such that $\fp$ is even, $c \,\GCD(2N,\fp)$ is not divisible by any $x$, where $x$ is an even number strictly greater than $2$ and $x$ divides $2N$, and $c$ is coprime to $\frac{2N}{\GCD(2N,\fp)}$.} Using the method of \cite{Closset:2021lwy}, we find that each theory in this subclass has a trivial 1-form symmetry. The 2-group symmetry is therefore, of course, trivial.
\end{enumerate}

\subsubsection{3d mirror theories}

We propose that the mirror theory for the $C^{2N}[p]$ theory with 1 mass parameter is described by the 3d $\CN=4$ $\U(1)$ gauge theory with $p-2N+1$ hypermultiplets of charge 1, along with 
\bes{
H_{\text{free}}= \frac{1}{2}(p-2N+1)(N-1)
}
free hypermultiplets.

For $N=1$, $C^{2}[p]$ is simply the $(A_1, A_{2p-3})$ theory, whose mirror is the $\U(1)$ gauge theory with $p-1$ flavors; in agreement with the above statement.  For $N=2$ (and so $p$ must be odd), the non-Higgsable SCFT is $(A_1, A_{p-3})$.  For $N=3$ and $p=3m$, the non-Higgsable SCFT is $C^6[2m]$, whose 1-form symmetry is $\BZ_2$.

The mirror theory of the $D_p(\USp(2N))$ theory with 1 mass parameter is therefore described by $H_{\text{free}}$ free hypermultiplets together with
\bes{ \label{mirrDpUSp2Nonemass}
B_0-C_1-B_1-C_2-\cdots-B_{N-1}\,-\,&C_{N}\,\begin{tikzpicture}[baseline, yshift=0.1cm, font=\scriptsize]  
\draw[blue] (0,0) to node[above]{$N$} (1,0);
\end{tikzpicture}
\U(1)-[p-2N+1] \\
&\,\,|\\
&\!\![B_{0}]
}
where the blue line denotes $N$ copies of the hypermultiplets in the bifundamental representation of $\USp(2N) \times \U(1)$.  Note that the Higgs branch symmetry $\SO(2N+1)$ of the $T[\SO(2N+1)]$ theory\footnote{As discussed below \eref{quivTSO2Mpp1}, the Coulomb branch symmetry of $T[\SO(2N+1)]$ theory is $\USp(2N)/\BZ_2$.} is decomposed into $\SO(2N) \times \O(1) \supset \SU(N) \times \U(1) \times \O(1)$, where the $\U(1)$ factor is gauged and is coupled to $p-2N+1$ hypermultiplets with charge 1.  Upon decoupling the tail $B_0-C_1-B_1-\cdots-C_N$, we obtain the SQED with $p-2N+1$ electrons, which is the mirror theory for $C^{2N}[p]$ as expected.

Let us consider the case of $N=1$, namely $D_p(\USp(2))$.  This theory turns out to be identical to the $(A_1, D_{2p})$ theory. In this special case, \eref{mirrDpUSp2Nonemass} gives the following mirror theory
\bes{ \label{mirrDpUSp2a}
B_0\,-\,&C_1-\U(1)-[p-1]\\
&\,\, | \\
&\!\![B_0]
}
It is worth comparing this theory with the known mirror theory of $(A_1, D_{2p})$ given by \cite[(4.17)]{Carta:2021whq}:
\bes{ \label{mirrDpUSp2b}
[1]-\U(1)-\U(1)-[p-1]
}
where an overall $\U(1)$ has been decoupled in \cite[(4.17)]{Carta:2021whq}.  To reconcile \eref{mirrDpUSp2a} with \eref{mirrDpUSp2b}, we first remark that, in \eref{mirrDpUSp2b}, we gauge the $\U(1)$ Cartan subalgebra of the flavor symmetry algebra $\su(2)$ of the $\U(1)$ gauge theory with 2 hypermultiplets of charge 1, which flows to the $T[\SU(2)]$ theory, and couple it to $p-1$ hypermultiplets. As pointed out in \cite{Gaiotto:2008ak}, the $T[\SU(2)]$ theory has another description as the $T[\SO(3)]$ theory, whose quiver description is $B_0 - C_1-[B_1]$. Gauging $\U(1)$ Cartan subalgebra of the flavor symmetry algebra $\so(3)$ of the latter theory and coupling it to $p-1$ hypermultiplets, we obtain \eref{mirrDpUSp2a}, as required.  This also provides a non-trivial test for the flavor node $[B_0]$, which arises as the commutant of such gauging, in \eref{mirrDpUSp2a}.

\subsection{Number of mass parameters greater than 1}
Let $\mu \geq 2$ be the number of mass parameters of the $D_p(\USp(2N))$ theories with $p>2N$.  We find that $p$ and $N$ have to satisfy the following conditions:
\bes{ \label{defparameters}
N &= \mu (2\fN-1)~, \quad \fN \in \BZ_{\geq 1}~, \qquad \text{and} \\ 
p &= 2\mu c~,~ \text{with $c$ coprime to $2\fN-1$}.
}
Note that $\GCD(p, 2N)=2\mu$ and $q \equiv \frac{2N}{\GCD(p, 2N)} = \frac{N}{\mu}= 2\fN-1$, which is odd. Due to the latter, this family of theories does not contain one with a 4d $\CN=2$ Lagrangian description.

The non-Higgsable SCFTs of these theories are
\bes{ \label{nHSCFTs1}
(A_{\fm-1},A_{2\fN-2})^{\otimes \mu}
}
where
\bes{
\fm \equiv \frac{p-2N}{\GCD(p, 2N)} = \frac{p-2N}{2\mu} = c - (2\fN-1)~.
}
It is worth remarking that the non-Higgsable sector \eref{nHSCFTs1} is identical to that of the $(A_{2\mu \fm-1}, D_{2\mu \fN-\mu+1})$ theory; see \cite[(6.17)]{Carta:2021whq}.

Having been identified the non-Higgsable sector, we can utilize the result of \cite{Carta:2022spy}, namely the 1-form symmetry of the $D_p(\USp(2N))$ in question must be the same as that of \eref{nHSCFTs1}. However, the latter is known to be trivial \cite{Closset:2020scj, DelZotto:2020esg, Hosseini:2021ged}, and so we conclude that the $D_p(\USp(2N))$ theories with higher mass parameters have a trivial 1-form symmetry and hence trivial 2-group symmetry.

\subsubsection{3d mirror theories} \label{sec:mirrormultmasses}
The mirror theory of $C^{2N}[p]$, with $\mu \geq 2$ mass parameters, is described by a quiver of $\mu+1$ $\U(1)$ gauge nodes such that $\mu$ of them form a complete graph with equal edge multiplicity $2 \fN \fm$, 
and the remaining $\U(1)$ node is connected to the others by the edges with equal multiplicity 
\bes{
F \equiv [\mu \fN - (\mu-2)] \fm +(2\fN-1)~.
}
Note that this quiver has an overall $\U(1)$ that decouples. There are also the following number of free hypermultiplets:
\bes{
H_{\text{free}} = \mu(\fN-1)(\fm-1)~,
}
equal to the total rank of the non-Higgsable SCFTs \eref{nHSCFTs1}.\footnote{The Higgs branch dimension of the mirror theory, excluding the free hypermultiplets, is $\frac{1}{2}\mu(\mu-1)(2 \fN \fm) + \mu F - \mu$. It can be checked that this quantity plus $H_{\text{free}}$ is equal to the rank of the corresponding $C^{2N}[p]$ theory in question.  Moreover, it can be checked that $\mu$ plus the total value of $24(c-a)$ of \eref{nHSCFTs1} is equal to $24(c-a)$ of the $C^{2N}[p]$ theory in question. These two tests provide a highly non-trivial test of our proposal.}

Upon decoupling the overall $\U(1)$, we obtain the aforementioned complete graph such that each $\U(1)$ gauge node has $F$ flavors of hypermultiplet of charge $1$.  The mirror of the corresponding $D_p(\USp(2N))$ theory is therefore described by the $T[\SO(2N+1)]$ tail, namely $B_0-C_1-B_1-\cdots-C_N$,\footnote{This tail gives rise to the CB symmetry $\USp(2N)$ in the IR.} such that the $C_N$ gauge node is connected to each of the $\mu$ $\U(1)$ nodes in the aforementioned flavored quiver with equal edge multiplicity $N/\mu = q = 2\fN-1$, and that there is a node $[B_0]$ attached to the $C_N$ gauge node, giving rise to a half-hypermultiplet in the latter. There are also $H_{\text{free}}$ hypermultiplets. 

\subsubsection*{Special case: \texorpdfstring{$2N$}{} divides \texorpdfstring{$p$}{}}
Let us consider the $D_p(\USp(2N))$ theories such that $2N$ divides $p$. These theories have $\GCD(p, 2N)/2=N$ mass parameters.  Here $\mu =N$, $\fN=1$, and $\fm = \frac{p}{2N}-1$.

The mirror theory of the $C^{2N}[p]$ theory is described by a quiver with $N+1$ $\U(1)$ gauge nodes: $N$ of them form a complete graph such that each of which edge has multiplicity $2\fm$ and the other $\U(1)$ gauge node is connected to the others by the edges, each with multiplicity $2\fm+1$. There is no free hypermultiplets in this case. Decoupling the overall $\U(1)$, we obtain the said complete graph such that each $\U(1)$ node has $2\fm+1$ flavors of hypermultiplets of charge $1$.

The mirror theory of the corresponding $D_p(\USp(2N))$ is the $T[\SO(2N+1)]$ tail, namely $B_0-C_1-B_1-\cdots-C_N$, such that the $C_{N}$ node is connected to each of the $N$ $\U(1)$ nodes in the aforementioned flavored quiver with edge multiplicity $1$, and that there is a node $[B_0]$ attached to the $C_N$ gauge node, giving rise to a half-hypermultiplet in the latter.

\subsubsection{Partially closing the full puncture of  \texorpdfstring{$D_{2N}(\USp(2N))$}{}} 
Let us now consider a set of models with multiple mass parameters. We focus on $D_p(\USp(2N))$ theories with $p=2N$ such that $N\geq 4$. This theory has $N$ mass parameters. In the notation of Section \ref{sec:mirrormultmasses}, we have $\fN=1$, $\mu=N$, $c=1$ and $\fm=0$. This is the smallest value of $p$ for which the regular puncture can be fully closed, leading to a free theory in the infrared.

We find a more interesting result by considering a partial closure, such as the puncture $[2N-2,2]$. In this case, the infrared theory turns out to be the following quiver with $\SU(2)\times \SU(3)^{N-3}$ gauge group: \bes{\label{ccsu3}
\begin{tikzpicture}[baseline]
\node[] (L1) at (0,0) {$\SU(2)$};
\node[] (L2) at (2,0) {$\SU(3)$};
\node (L7) at (2,1.2) {$[1]$};
\node[] (L3) at (4,0) {$\dots$};
\node[] (L4) at (6,0) {$\SU(3)$};
\node[] (L5) at (8,-1) {$(A_1,D_4)$};
\node[] (L6) at (8,1) {$(A_1,D_4)$};
\draw[thick]  (L1) to (L2) to (L3) to (L4) to (L5); 
\draw[thick]  (L4) to (L6);
\draw[thick]  (L2) to (L7);
\end{tikzpicture}
}
where $(A_1,D_4)$ denotes the rank-1 AD theory whose Higgs branch coincides with the reduced one $\SU(3)$ instanton moduli space. As in the previous cases, we can replace the two copies of the $(A_1, D_4)$ theory with a single copy of a rank-2 theory whose Higgs branch is the reduced two $\SU(3)$ instanton moduli space. The quiver \eqref{ccsu3} with such a rank-2 theory on the right arises if we close the regular puncture to $[2N-2,1^2]$. Again, by giving an expectation value to the $\SU(2)$ moment map of the latter theory we recover \eqref{ccsu3} and this indeed corresponds to closing the puncture back to $[2N-2,2]$. 

This observation allows us to test our proposal in Section \ref{sec:mirrormultmasses} for the 3d mirror theory of $D_p(\USp(2N))$. In fact, the 3d mirror of \eqref{ccsu3} can be determined and turns out to be 
\bes{ \label{mirrD2NUSp2Ncloseto}
\begin{tikzpicture}[baseline,scale=1.2]
\node[] (a) at (-0.5,0) {$B_0$};
\node[] (b) at (1,0) {$C_1$};
\node (c) at (1,1) {$[B_0]$};
\node[] (d) at (2.5,1) {$\U(1)$}; 
\node (e) at (4,1) {$[1]$};
\node[] (f) at (2.5,-1) {$\U(1)$}; 
\node (g) at (4,-1) {$[1]$}; 
\node[] (h) at (2.5,0.2) {$\U(1)$}; 
\node[] at (2.5,-0.3) {$\vdots$};
\node (l) at (4,0.2) {$[1]$};
\draw (a)--(b) (b)--(c) (b)--(d) (d)--(e) (b)--(f) (f)--(g) (b)--(h) (h)--(l);
% \draw[snake=brace]  (4.5,1.3) -- (4.5,-1.3);
%   \node[] at (5,0) {$N$};
   \draw [line width=1pt,decorate, decoration = {brace}] (4.5,1.3) --  node[right,pos=0.5,xshift=0.2cm] {$N$} (4.5,-1.3);
\end{tikzpicture}
}
In order to reconcile this with the proposal in Section \ref{sec:mirrormultmasses}, we recall that the mirror theory for $D_{2N}(\USp(2N))$ can be realized as a quiver of $N$ $\U(1)$ nodes such that each node has $1$ flavor of hypermultiplet of charge $1$, and each of the $N$ $\U(1)$ node is connected to the $C_N$ node in the following tail
\bes{
B_0 - C_1 - B_1 - \cdots - &C_N \\
& \,\, | \\
& \!\! [B_0]
}
where it should be noted that the theory associated with the $C$-partition $[1^{2N}]$ is $B_0-C_1-B_1-\cdots-[B_N]$. If, on the other hand, one considers the $C$-partition $[2N-2,2]$, the quiver gets ``shortened'' to $B_0-C_1-[B_{N}]$.  Therefore, partially closing the full puncture $[1^{2N}]$ in the original 4d theory to $[2N-2]$ amounts to shortening the above tail to
\bes{
B_0 - &C_1 \\
& \,\, | \\
& \!\! [B_0]
}
Attaching the $C_1$ node to the aforementioned $N$ $\U(1)$ nodes, one obtains \eref{mirrD2NUSp2Ncloseto} as required.

\acknowledgments

N.~M. would like to express his gratitude to a number of people for several useful discussions: Fabio Apruzzi, Pietro Benetti Genolini, Cyril Closset, Ho Tat Lam, Matteo Sacchi, Shu-Heng Shao, Alberto Zaffaroni, and Gabi Zafrir. He gratefully acknowledges support from the Simons Center for Geometry and Physics, Stony Brook University, at which this project is completed. F. C. is supported by STFC consolidated grant ST/T000708/1. The work of S. G. is supported by the INFN grant ``Per attività di formazione per sostenere progetti di ricerca'' (GRANT 73/STRONGQFT). A. M. is supported in part by Deutsche Forschungsgemeinschaft under Germany's Excellence Strategy EXC 2121  Quantum Universe 390833306. 

\appendix

\section{Absence of the 2-group structure for other \texorpdfstring{$D_p(G)$}{} theories}
\label{sec:abseDpG}

In this appendix, we discuss the absence of $2$-group structure for other $D^b_p(G)$ theories than $D_p(\USp(2N))$.  For example, the $D^b_p(\SU(N))$ theories do not have a $2$-group structure, since they have a trivial $1$-form symmetry \cite{Hosseini:2021ged}. One might wonder if this holds in other $D_p(G)$ theories, for which there is a non-trivial $1$-form symmetry. The purpose of this appendix is to show that there is no $2$-group structure for $D_p(\SO(2N))$, $D_p^{N+1}(\USp(2N))$, $D^b_p(\USp'(2N))$ and $D^b_p(\SO(2N+1))$. We thus see that $D_p(\USp(2N))$ is special among the $D_p(G)$ theories when it comes to the presence of the 2-group structure.  

\subsection{\texorpdfstring{$D_{p}(\SO(2N))$}{} theories}
\label{sec:abseDpSO}
Using the results of \cite{Giacomelli:2017ckh, DelZotto:2020esg, Hosseini:2021ged}, it can be deduced that any $D_p(\SO(2N))$ theory with a non-trivial 1-form symmetry must have either zero mass parameter or one mass parameter.\footnote{As mentioned earlier, we do not count the mass parameters associated with the $\SO(2N)$ flavor symmetry.} However, it should be emphasized that not all theories with zero or one mass parameter have a non-trivial 1-form symmetry. Since any theory with a non-trivial 2-group structure must have a non-trivial 1-form symmetry, we will analyze these two classes of theories in the following.

\subsubsection*{One mass parameter} 
We consider the theories with one mass parameter. Let us start from a Lagrangian theory, namely the $D_p(\SO(2Mp+2))$ theory, with {\it $p$ even} (see \cite[Appendix C.2]{Cecotti:2013lda} and \cite[(8.32)]{Carta:2021whq}):
\bes{
\scalebox{0.9}{$
\begin{split} \label{LagDpSOa}
[D_{Mp+1}]-C_{Mp-M}-D_{Mp-2M+1}-C_{Mp-3M}-D_{Mp-4M+1}-\cdots-D_{2M+1}-C_M-[D_{1}]
\end{split}$}
}
Suppose that we take each $D_n$ gauge group to be $\Spin(2n)$. Then,  this theory has an electric $\BZ_2^{p/2-1}$ 1-form symmetry. Moreover, this theory has the operators that transform under the representation $[\mathbf{2Mp+1}; \pm 1]$ of $\so(2Mp+2) \times \mathfrak{u}(1)$, and so the global form of the flavor symmetry is $\SO(2Mp+2) \times \U(1)$, and not $\frac{\SO(2Mp+2) \times \U(1)}{\BZ_2}$.  Since the quotient of the flavor symmetry is trivial, there is no obstruction class analogous to the 2nd Stiefel-Whitney class, and so there is no 2-group structure. 

The other theories with one mass parameters can be obtained from the $D_p(\SO(2Mp+2))$ theory by bootstrapping, namely by considering $D_{\fp c}(\SO(2M\fp+2))$ theories with $c$ coprime to $q \equiv \frac{2M \fp}{\GCD(\fp, 2M \fp)} = 2M$.  Since this process preserves the 1-form symmetry and the global form of the flavor symmetry, we conclude that every $D_p(\SO(2N))$ with one mass parameter has a trivial 2-group symmetry.

\subsubsection*{Zero mass parameter} 
As before, let us start with the theories with a Lagrangian description, namely $D_p(\SO(2Mp+2))$, with {\it $p$ odd} (see \cite[Appendix C.2]{Cecotti:2013lda} and \cite[(8.32)]{Carta:2021whq}):
\bes{ \label{LagDpSO}
[D_{Mp+1}]-C_{Mp-M}-D_{Mp-2M+1}-C_{Mp-3M}-D_{Mp-4M+1}-\cdots-C_{2M}-D_{M+1}
}
Suppose that we take each $D_n$ gauge group to be $\Spin(2n)$. This theory then has an electric $\BZ_2^{(p-1)/2}$ 1-form symmetry.  The Higgs branch is generated by the moment map operator in the adjoint representation of $\so(2Mp+2)$ flavor symmetry algebra, and so the global form of the flavor symmetry is $\SO(2Mp+2)/\BZ_2$.  If $M$ is odd, $Mp$ is odd, and so all the $D$-type gauge groups are of even rank; in which case, there is no two-group structure \cite{Lee:2021crt}.  If $M$ is even, $Mp$ is even, all $D$-type gauge groups are of odd rank, and the global form of the flavor symmetry can be written as $\SO(2Mp+2)/\BZ_2= \Spin(2Mp+2)/\BZ_4$. We propose that there is also no 2-group structure in this case.  This can be seen, for example, by taking $p=3$ and $M=2$, where the quiver becomes $[\Spin(7)/\BZ_4]-\USp(8)-\Spin(6)$, and observing that there is no non-trivial extension between the $\BZ_2$ 1-form symmetry and the $\BZ_4$ quotient in the flavor symmetry (\eg the maximal trivially acting group is not $\BZ_8$). In fact, as shown in \cite[(8.40)]{Bhardwaj:2021mzl}, for general even $M$ and odd $p$, the corresponding short exact sequence in $0 \rightarrow \BZ_2^{(p-1)/2} \rightarrow \BZ_2^{(p-1)/2} \times \BZ_4 \rightarrow \BZ_4 \rightarrow 0$ splits, and so there is no 2-group structure. 

The other theories with zero mass parameter can be obtained by bootstrapping the $D_p(\SO(2Mp+2))$ theory with $p$ odd.  In other words, they are of the form $D_{\fp c}(\SO(2M\fp+2))$ with $c$ coprime to $q \equiv \frac{2M\fp}{\GCD(2M\fp, \fp)} = 2M$. As discussed in \cite{Bhardwaj:2021mzl}, each theory has a $\BZ_2^{(\fp-1)/2}$ 1-form symmetry, as for the case of $c=1$. Moreover, using the mirror theory described by \cite[(5.4)]{Carta:2021whq}, we see that the global form of the symmetry is $\SO(2M\fp+2)/\BZ_2$, precisely as for the case of $c=1$.  The relevant exact sequence for these theories is still given by \cite[(8.40)]{Bhardwaj:2021mzl}, and so there is no 2-group structure for $c\geq 1$. 

Finally, let us comment on the case of \cite[(8.42)-(8.43)]{Bhardwaj:2021mzl}, where it is claimed that there is a non-trivial extension between the 1-form symmetry (denoted by $H^T_\CP$) and the center of the flavor symmetry (denoted by $Z_\CP$), and so it is possible that the theories discussed there possess a 2-group symmetry. Since the center of the flavor symmetry is $Z_\CP = \BZ_2 \times \BZ_2$, its symmetry algebra is of the $D$-type with even rank.  Suppose that one of the theories in this class admits a quiver description, where every gauge group has zero-beta function. Then, either all the $D$-type groups in the quiver (including the flavor node) have odd rank or either all of them have even rank.  Since we knew that the flavor symmetry is of the $D$-type of even rank, all gauge groups of the $D$-type must have even rank as well. Therefore, if we assume that the quiver theory in question is conformal, it cannot have a 2-group structure and no non-trivial extension between the 1-form symmetry and the quotient of the flavor symmetry. We thus conclude that the conformal quiver does not belong to the case of \cite[(8.42)-(8.43)]{Bhardwaj:2021mzl}. In other words, for the Lagrangian theories, the case of \cite[(8.42)-(8.43)]{Bhardwaj:2021mzl} corresponds to asymptotically free theories but not conformal theories.

\subsection{\texorpdfstring{$D_{p}^{N+1}(\USp(2N))$}{} theories} 
\label{sec:abseDpUSp}

We recall from \cite[(2.15)]{Carta:2021whq}, that there is a second series of $D_p(\USp)$ models dubbed $D_p^{N+1}(\USp(2N))$ which is described by the $\mathbb{C}^3\times \mathbb{C}^*$ hypersurface in Type IIB 
\be\label{defsing}
u^2+x^N+xy^2+yz^p+\text{defs.}=0
\ee 
where $p$ is \textit{half-odd-integer} and $z$ is the $\mathbb{C}^*$ variable. 
We claim that, unlike the $D_p(\USp(2N))$ series, which we have discussed in the main body of the paper, this class of models does not exhibit any non-trivial 2-group structure. 

We can start by observing that $D_p^{N+1}(\USp(2N))$ do not exhibit any mass parameter apart from those related to the $\USp(2N)$ global symmetry. Another key fact which drastically simplifies the analysis is that via the bootstrapping procedure, we can always restrict the analysis to the Lagrangian subclass. 

The easiest way to proceed is to notice that from \eqref{defsing} we can conclude that the spectral equation for the Hitchin field $\Phi$ of $D_p^{N+1}(\USp(2N))$ theories reads (see e.g. \cite{Giacomelli:2017ckh})
\be \Phi^{2N+2}+z^{2p}+\text{defs.}=0\ee
and therefore the monodromy of the Hitchin field is fully encoded in the quantity 
\begin{equation}
    q=\frac{2N+2}{\GCD(2p,2N+2)}\coma
    \label{eq:qUSp2}
\end{equation}
which does not change if we replace $2p$ with $\GCD(2p,2N+2)$. In the latter case, the corresponding SCFT is the linear quiver 
\be \USp(q-2)-\Spin(2q)-\USp(3q-2)-\dots-\Spin(2N+2-q)-[\USp(2N)]\ee 
Notice that the quantity $q$ is always an even integer and therefore all the $D$-type groups in the linear quiver have even rank and therefore their center is $\mathbb{Z}_2^2$ and not $\mathbb{Z}_4$. We conclude that the short exact sequence in this case always splits, and the 2-group structure is trivial \cite{Lee:2021crt, Apruzzi:2021mlh}.

\subsection{\texorpdfstring{$D_p(G)$}{} theories of the twisted \texorpdfstring{$A$}{}-type} \label{sec:twistedA}

Theories in this class are defined by compactifying the 6d theory of type $A_{n-1}$ on a sphere with two twisted punctures, one regular and one irregular. The relevant hypersurface singularities in $\mathbb{C}^3\times \mathbb{C}^*$ are of the form 
\be u^2+x^2+y^{n}+ z^p + \text{defs.} = 0\coma\ee 
or 
\be u^2+x^2+y^{n}+ y z^p + \text{defs.} = 0.\ee 
Since the punctures are twisted, the Casimir invariants of odd degree (included inside defs. in the equations above) are proportional to a polynomial in $z$ times $\sqrt{z}$, whereas the Casimirs of even degree are just polynomials in $z$. All the deformations of this type describe VEV of CB operators, mass parameters, relevant and marginal couplings. Clearly, the only difference between the theories described by the hypersurfaces
\be\label{case1} u^2+x^2+y^{n}+ z^p + \text{defs.} = 0\coma\ee 
and 
\be\label{case2} u^2+x^2+y^{n+1}+ y z^p + \text{defs.} = 0.\ee 
lies in the fact that the latter includes all the parameters of the former plus the Casimir of degree $n+1$. Since we are interested in determining the 2-group structure, it suffices to identify Lagrangian theories and for this purpose \eqref{case1} and \eqref{case2} can be analyzed together. We therefore have to discuss separately two cases: \eqref{case1} with $n$ even and with $n$ odd. The parity of $n$ leads, upon twisting, to SCFTs with different global symmetry: $\SO(n+1)$ when $n = 2m$, and $\USp(n-1)$ when $n=2m+1$, for $m\in \mathbb{Z}_{\geq 1}$. In \cite{Carta:2021whq}, these models were dubbed $D_p(\SO(n+1))=D_p(\SO(2m+1))$ and $D_p(\USp'(n-1)) =D_p(\USp'(2m))$ respectively. However, for consistency of notation and clarity, here, we call them $D_p^n(\SO(n+1))= D
^{2m}_p(\SO(2m+1)) $ and $D_p^{n}(\USp'(n-1)) = D_p^{2m+1}(\USp'(2m))$.\footnote{On the other hand, \eref{case2}, upon twisting, leads to $\SO(n+2)$ when $n=2m-1$, and $\USp(n)$ when $n=2m$, with $m\in \mathbb{Z}_{\geq 1}$.  Thus, it describes $D
^{2m-1}_p(\SO(2m+1))$ and $D_p^{2m}(\USp'(2m))$, for $n=2m-1$ and $n=2m$, respectively.}  The result is that the 2-group structure is trivial for all these models and the Lagrangian subclass is given by a linear quiver with alternating gauge groups (all with vanishing beta function) of $B$ and $C$ types, with a flavor node only at one end of the quiver \cite{Lee:2021crt, Apruzzi:2021mlh}. Explicitly, the Lagrangian $D_p^{2m}(\USp'(2m))$ theories (with zero mass parameter) can be obtained by taking $m=2p\mathfrak{m}-\left(p+\frac{1}{2}\right)$,\footnote{Recall that $p$ is half-odd-integer.} with $\mathfrak{m}\in \ZZ_{\geq 1}$:
\bes{
\begin{split}
B_0 - C_{\mathfrak{m}-1}- B_{2\mathfrak{m}-1} &- C_{3\mathfrak{m}-2}- B_{4\mathfrak{m}-2} - C_{5\mathfrak{m}-3} - \cdots \\&\cdots - C_{(2 p-2)\fm -\left(p-\frac{1}{2}\right)} - B_{(2 p-1)\fm -\left(p-\frac{1}{2}\right)}- \left[C_{2p\mathfrak{m}-\left(p+\frac{1}{2}\right)}\right] \end{split}
}
whereas the Lagrangian $D_p^{2m}(\SO(2m+1))$ theories (with zero mass parameter) can be obtained by taking $m = 2p\mathfrak{m}-p$, with $\mathfrak{m}\in \ZZ_{\geq 1}$:
\bes{
\begin{split}
B_0 - C_{\mathfrak{m}-1}- B_{2\mathfrak{m}-1} &- C_{3\mathfrak{m}-2}- B_{4\mathfrak{m}-2} - C_{5\mathfrak{m}-3} - \cdots \\ &\cdots - B_{(2 p-1)\fm -(p-1)}- C_{(2 p-1)\fm -p} - \left[B_{2p\mathfrak{m}-p}\right]~.\end{split}
}

\section{A comment on the Higgs branch of bootstrapped theories}
\label{sec:HBbootstrap}
Theories in the same bootstrap family have very similar properties. One then can ask if also their Higgs branches have a common structure. In this appendix, we observe that for every two $D_p(\SU(N))=D_{\mu c}(\SU(q\mu))$ theories with $p\geq N$, e.g. $\mathcal{T}_1$ and $\mathcal{T}_2$, linked by bootstrap as explained in \cref{sec:bootstrap,sec:bootstrapmirror} the following holds:
\begin{enumerate}
    \item The dimension of the Higgs branch of $\mathcal{T}_1$ is equal to that of $\mathcal{T}_2$. This fact is clear from the magnetic quivers, as all the magnetic quivers of the same family of bootstrap theories have the same rank, and just differ by the number of bifundamental hypermultiplets, or of the free hypermultiplets, as in \eqref{eq:DpSUhyper}.
    \item The Hasse diagram \cite{Bourget:2019aer}, encoding the foliation structure of symplectic leaves and transverse slices, has the same shape both for $\mathcal{T}_1$ and $\mathcal{T}_2$. Here with \emph{same shape} we mean that we consider the two Hasse diagrams just as unoriented graphs, disregarding the labels attached to the various lines. When such labels are disregarded, then the two unoriented graphs coincide.
    \item Suppose now that the bootstrap shift parameter for $\mathcal{T}_2$ is larger than the one for $\mathcal{T}_1$, then each of the transverse slices of the Higgs branch of $\mathcal{T}_2$ belongs to the same family of singularities of the corresponding slice of the Higgs branch of $\mathcal{T}_1$, but of a higher rank.\footnote{Here we use the word \emph{rank of a singularity} to denote the number of exceptional $\mathbb{P}^1$s needed to perform a completely smooth small resolution of the singularities. For instance, we would say that the Kleinian singularity $\mathbb{C}^2/\mathbb{Z}_2$ has rank $1$, while $\mathbb{C}^2/\mathbb{Z}_{10}$ has rank $9$.}
\end{enumerate}
We observe that this phenomenon is preserved also after closing the full puncture, such that the $D_p(\SU(N))$ theory flows to the  $(A_{p-N-1},A_{N-1})$ theory. To illustrate this observation, we decided to consider the Hasse diagrams for the AD theories $(A_n,A_m)$ discussed in \cite{Bourget:2019aer}. In particular, let us consider the case in which $p$ is multiple of $N$, such that we can discuss the Hasse diagrams of the 3d mirror theories of $(A_{\mu(c-1)-1},A_{\mu-1})$ in \cref{tab:hassecompletegraphs2}. For any fixed complete graph with $\mu$ $\U(1)$ nodes, the Hasse diagram will look the same, but with different transverse slices whose associated singularities change with the bootstrap parameter $c$.

 \begin{table}[!htp]
 	\centering
 	\begin{tabular}{|m{1.3cm}|m{2.4cm}|m{5cm}|}
 	\hline
 	$\mu=2$ & $\mu=3$ & $\mu=4$ 
 	 \\ \hline
 	 	\begin{tikzpicture}[node distance=30pt]
 	\tikzstyle{hasse} = [circle, fill,inner sep=1pt];
 		\node at (-0.7,-0.5) [] (1a) [] {};
 		\node at (0,-0.5) [hasse] (1b) [label=right:\footnotesize{$1\;\;$}] {};
 	           \node   (0b) [above of=1b] {};
 		\node at (0.7,-0.5) [] (1c) [] {};
 		\node [] (2a) [below of=1a] {};
 	        \node  (3a) [below of=2a] {};%
 		\node  (4a) [below of=3a] {};%
 		\node [hasse] (2b) [label=right:\footnotesize{$0$},below of=1b] {};
 		\draw (1b) edge [] node[label=center:\footnotesize{$A_{c-2}\;\;\;\;\;\;\;\;\;$}] {} (2b);
 	\end{tikzpicture}
 	 &
 	 	\begin{tikzpicture}[node distance=30pt]
 	\tikzstyle{hasse} = [circle, fill,inner sep=1pt];
 		\node at (-0.7,-0.5) [] (1a) [] {};
 		\node at (0,-0.5) [hasse] (1b) [label=above:\footnotesize{$2$}] {};
 		\node at (0.7,-0.5) [] (1c) [] {};
 \node [hasse] (2a) [below of=1a] {};
 \node [hasse] (2b) [below of=1b] {};
 \node [hasse] (2c) [label=right:\footnotesize{$1$},below of=1c] {};
 \node [] (3a) [below of=2a] {};
 \node [hasse] (3b) [label=right:\footnotesize{$0$},below of=2b] {};
 \node [] (3c) [below of=2c] {};
 \draw 
 (1b) edge [] node[label=center:\footnotesize{$A_{c-2}\;\;\;\;\;\;\;\;\;\;\;$}] {} (2a)
 (2a) edge [] node[label=center:\footnotesize{$A_{2c-3}\;\;\;\;\;\;\;\;\;\;$}] {} (3b)
 (1b) edge [] node[] {} (2b)
 (2b) edge [] node[] {} (3b)
 (1b) edge [] node[] {} (2c)
 (2c) edge [] node[] {} (3b);
 	\end{tikzpicture}
 	 &
 	\begin{tikzpicture}[node distance=30pt]
 	\tikzstyle{hasse} = [circle, fill,inner sep=1pt];
 		\node at (-1.2,-0.5) [] (1a) [] {};
 		\node at (-0.8,-0.5) [] (1b) [] {};
 		\node at (-0.4,-0.5) [] (1c) [] {};
 		\node at (0,-0.5) [hasse] (1d) [label=above:\footnotesize{$3$}] {};
 		\node at (0.4,-0.5) [] (1e) [] {};	
 \node at (0.8,-0.5) [] (1f) [] {};
 \node at (1.2,-0.5) [] (1g) [] {};
 \node [hasse] (2a) [below of=1a] {};
 \node [hasse] (2b) [below of=1b] {};
 \node [hasse] (2c) [below of=1c] {};
 \node [] (2d) [below of=1d] {};
 \node [hasse] (2e) [below of=1e] {};
 \node [hasse] (2f) [below of=1f] {};
 \node [hasse] (2g) [label=right:\footnotesize{$2$},below of=1g] {};
 \node [hasse] (3a) [below of=2a] {};
 \node [hasse] (3b) [below of=2b] {};
 \node [hasse] (3c) [below of=2c] {};
 \node [hasse] (3d) [below of=2d] {};
 \node [hasse] (3e) [below of=2e] {};
 \node [hasse] (3f) [below of=2f] {};
 \node [hasse] (3g) [label=right:\footnotesize{$1$},below of=2g] {};
 \node [] (4d) [below of=3d] {};
 \node [hasse] (5d) [label=right:\footnotesize{$0$},below of=4d] {};
 \draw 
 (2a) edge [] node[label=left:\footnotesize{$A_{c-2}$}] {} (3a)
 (2a) edge [orange] node[] {} (3d)
 (2a) edge [orange] node[] {} (3e)
 (2b) edge [] node[] {} (3a)
 (2b) edge [orange] node[] {} (3f)
 (2b) edge [orange] node[] {} (3g)
 (2c) edge [] node[] {} (3b)
 (2c) edge [orange] node[] {} (3d)
 (2c) edge [orange] node[] {} (3f)
 (2e) edge [] node[] {} (3b)
 (2e) edge [orange] node[] {} (3e)
 (2e) edge [orange] node[] {} (3g)
 (2f) edge [] node[] {} (3c)
 (2f) edge [orange] node[] {} (3d)
 (2f) edge [orange] node[] {} (3g)
 (2g) edge [] node[] {} (3c)
 (2g) edge [orange] node[] {} (3e)
 (2g) edge [orange] node[label=right:\footnotesize{$\;\;A_{2c-3}$}] {} (3f)
 (1d) edge [] node[label=above:\footnotesize{$A_{c-2}\;\;$}] {} (2a)
 (1d) edge [] node[] {} (2b)
  (1d) edge [] node[] {} (2c)
   (1d) edge [] node[] {} (2e)
  (1d) edge [] node[] {} (2f)
  (1d) edge [] node[] {} (2g)
   (3b) edge [red] node[] {} (5d)
   (3c) edge [red] node[] {} (5d)
   (3d) edge [green] node[] {} (5d)
   (3e) edge [green] node[] {} (5d)
   (3f) edge [green] node[] {} (5d)
   (3g) edge [green] node[label=right:\footnotesize{$A_{3c-4}$}] {} (5d)
     (3a) edge [red] node[label=left:\footnotesize{$A_{4c-5}$}] {} (5d);
 	\end{tikzpicture}
 	\\ \hline
 		\end{tabular}
 	\caption{Hasse diagrams taken and adapted from \cite[Table 10]{Bourget:2019aer} for the 3d mirror of $(A_{\mu(c-1)-1},A_{\mu-1})$ with $\mu=2,3,4$ and $c> 2$. The case for $c=2$ is in \cite[Table 9]{Bourget:2019aer}.
 	}\label{tab:hassecompletegraphs2}
 \end{table}
 
 We conjecture that such observation is true for any $D_p(G)$ for $p\geq h(G)$. It would be interesting to check if this is true also for theories whose magnetic quivers involve both unitary and orthosymplectic nodes. 

\bibliographystyle{JHEP}
\bibliography{mybib}

\end{document}